\documentclass[%
 reprint,
 amsmath,
 amssymb,
aps,
prb,
eprint,
]{revtex4-2}

\usepackage{graphicx}
\usepackage{dcolumn}
\usepackage{bm}
\usepackage{mathtools}
\usepackage{amsmath}
\usepackage{amssymb}
\usepackage{physics}
\usepackage{lipsum}
\usepackage{color}

\begin{document}

\preprint{APS/123-QED}

\title{Haldane model on the Sierpi\'nski gasket}

\author{Z.~F.~Osseweijer}
\email{z.f.osseweijer@students.uu.nl}
\affiliation{Institute of Theoretical Physics, Utrecht University, Utrecht, 3584 CC, Netherlands}
\author{L. Eek}
\affiliation{Institute of Theoretical Physics, Utrecht University, Utrecht, 3584 CC, Netherlands}
\author{A. Moustaj}
\affiliation{Institute of Theoretical Physics, Utrecht University, Utrecht, 3584 CC, Netherlands}
\author{M. Fremling}
\affiliation{Institute of Theoretical Physics, Utrecht University, Utrecht, 3584 CC, Netherlands}
\author{C. Morais Smith}
\affiliation{Institute of Theoretical Physics, Utrecht University, Utrecht, 3584 CC, Netherlands}


\date{\today}

\begin{abstract}
We investigate the topological phases of the Haldane model on the Sierpi\'nski gasket. 
As a consequence of the fractal geometry, multiple fractal gaps arise. Additionally, a flat band appears, and due to a complex next-nearest-neighbor hopping, this band splits and multiple topological flux-induced gaps emerge. 
Owing to the fractal nature of the model, conventional momentum-space topological invariants cannot be used. Therefore, we characterize the system's topology in terms of a real-space Chern number. In addition, we verify the robustness of the topological states to disorder.
Finally, we present phase diagrams for both a fractal gap and a flux-induced gap. Previous work on a similar system claims that fractality ``squeezes'' the well-known Haldane phase diagram.  
However, this result arises because a doubled system was considered with two Sierpi\'nski gaskets glued together. We consider only a single copy of the Sierpi\'nski gasket, keeping global self-similarity.
In contrast with these previous results, we find intricate and complex patterns in the phase diagram of this single fractal. 
Our work shows that the fractality of the model greatly influences the phase space of these structures, and can drive topological phases in the multitude of fractal and flux-induced gaps, providing a richer platform than a conventional integer dimensional geometry.
\end{abstract}

\maketitle

\section{\label{sec:intro}          Introduction}
For a long time, topological phases of matter have been at the forefront of condensed-matter research. This is because of their distinct and exotic properties, such as metallic boundary modes in an otherwise insulating bulk \cite{hatsugaiChernNumberEdge1993, hasanColloquiumTopologicalInsulators2010}. The first type of topological state, the quantum Hall effect, was discovered by von Klitzing in 1980 by applying a magnetic field perpendicular to a two-dimensional (2D) electron gas and performing transport measurements at low temperatures \cite{klitzingNewMethodHighAccuracy1980}.
In 1988, Haldane showed that the requirement of a strong magnetic field to observe the quantum Hall effect in 2D systems could be relaxed to a requirement of broken time-reversal (TR) symmetry \cite{haldaneModelQuantumHall1988}. In 2005, this idea was generalized by Kane and Mele with the introduction of spin-orbit coupling (SOC) in graphene, leading to the quantum spin Hall effect. 
In their seminal work, they showed that SOC breaks TR symmetry for a single spin, but that this symmetry gets restored by the other spin. Theoretically, this opens a topological gap in graphene \cite{kaneQuantumSpinHall2005, kaneTopologicalOrderQuantum2005}. 
Although they overestimated the strength of this effect, these results kick-started the field of topological insulators. Since then, topological phases have been extensively studied theoretically and experimentally in photonic \cite{rechtsmanPhotonicFloquetTopological2013, segevPhotonicTopologicalInsulators2014, wuSchemeAchievingTopological2015}, acoustic \cite{khanikaevTopologicallyRobustSound2015, gaoTopologicalSoundPumping2020}, and electronic systems \cite{kaneQuantumSpinHall2005, wanTopologicalSemimetalFermiarc2011, huTransportTopologicalSemimetals2019}. 
These systems fit in the `tenfold way' classification, which predicts the existence of topological phases based solely on the spectral symmetries and dimension of a system \cite{ryuTopologicalInsulatorsSuperconductors2010}. 

More recently, this framework has been expanded to describe non-Hermitian systems \cite{leykamEdgeModesDegeneracies2017, ashidaNonHermitianPhysics2020, liuTopologicalPhasesNonHermitian2023, eekEmergentNonHermitianModels2024}, non-equilibrium systems \cite{chernyakNonEquilibriumThermodynamicsTopology2009, dehghaniOutofequilibriumElectronsHall2015, schulerTracingNonequilibriumTopological2017, rudnerBandStructureEngineering2020}, quasicrystals \cite{krausTopologicalStatesAdiabatic2012, bandresTopologicalPhotonicQuasicrystals2016, fanTopologicalStatesQuasicrystals2022, moustajAnomalousPolarizationOnedimensional2024} and crystalline topology \cite{fuTopologicalCrystallineInsulators2011, hsiehTopologicalCrystallineInsulators2012, benalcazarQuantizationFractionalCorner2019, eekHigherorderTopologyProtected2024}, to cite only a few.  
A limitation of the `tenfold way' is that it only accounts for integer dimensions. 
However, it is known that fractals can have a non-integer Hausdorff dimension $d_H$ \cite{hausdorffDimensionUndAeusseres1918}, and host a new type of symmetry: self-similarity, which means that a subset of the system is similar to the entire system. In its most extreme case, self-similarity becomes scale invariance, when the subset and the entire system are indistinguishable. Some of the earliest works studying physical properties of fractals considered phase transitions \cite{gefen_critical_1980, gefen_solvable_1981, domany_solutions_1983, melrose_hierarchical_1983, gefen_phase_1983}.

\begin{figure}
    \includegraphics[width=\linewidth]{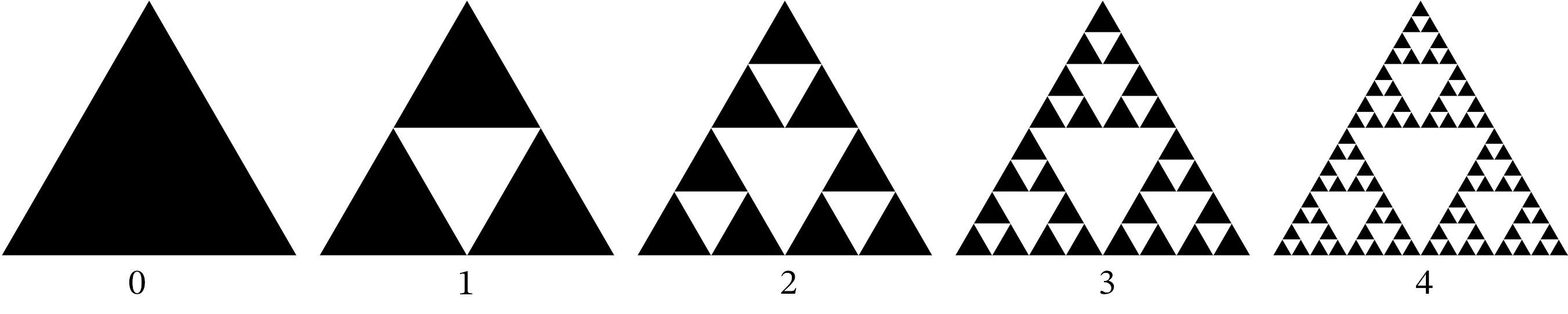}
    \caption{The first (zeroth to fourth) generations of a Sierpi\'nski gasket. Each generation contains three copies of the previous one, scaled down by a factor of two. Therefore, the Hausdorff dimension is given by $d_H = \log{3}/\log{2} \approx 1.585$. }
    \label{fig:SG}
\end{figure}

Two of the most commonly considered fractals are the Sierpi\'nski carpet and the Sierpi\'nski gasket \cite{gefen_critical_1980, newkomeNanoassemblyFractalPolymer2006, shangAssemblingMolecularSierpinski2015, vanveenQuantumTransportSierpinski2016, kempkesDesignCharacterizationElectrons2019, iliasovHallConductivitySierpinski2020, yangConfinedElectronsEffective2020, yangElectronicPropertiesQuantum2022, gefen_phase_1984, gefen_phase_1984-1}. In Fig.~\ref{fig:SG}, the Sierpi\'nski gasket is shown up to its fourth generation.
In 2015, Shang et al.~synthesized the first molecular realization of this fractal \cite{shangAssemblingMolecularSierpinski2015}. A few years later, Kempkes et al.~presented the first electronic quantum fractal using artificial atoms and showed that the local density of states followed the Hausdorff dimension of this fractal \cite{kempkesDesignCharacterizationElectrons2019}. Later, the dynamics in a photonic Sierpi\'nski gasket was studied and it was found that the diffusion exponent became the Hausdorff dimension after the photons met the first void 
of the fractal \cite{xuQuantumTransportFractal2021}. Very recently, a naturally occurring Sierpi\'nski gasket-shaped enzyme was observed \cite{sendkerEmergenceFractalGeometries2024}. 

In the last few years, these fractals and their possibly non-integer dimensionality have been combined with the study of topology, as an expansion of the `tenfold way'.
On the theoretical side, the (integer) quantum Hall effect on the Sierpi\'nski gasket has been investigated by M.~Brezezi\'nska et al.~in 2018 \cite{brzezinskaTopologySierpinskiHofstadterProblem2018} and further expanded upon in Refs.~\cite{fremlingExistenceRobustEdge2020, sarangi_adiabatic_2023, fischerRobustnessChiralEdge2021, pedersen_graphene_2020}. 
In 2020, the first discussion of the fractional quantum Hall effect was presented and the existence of anyons in these fractals was predicted \cite{mannaAnyonsFractionalQuantum2020}. 
Not much later, superconductivity, higher-order topology and non-Hermitian topology were theoretically investigated in a Sierpi\'nski gasket \cite{mannaHigherorderTopologicalPhases2022, manna_noncrystalline_2024, manna_inner_2023}. 
Finally, in 2023, a Hubbard model was considered on a Sierpi\'nski gasket and the effects of SOC and electron-electron interactions were revealed \cite{conteFractalLatticeHubbardModel2023}. 

The topological properties of fractals have also been investigated experimentally. In 2018, 
photonic Floquet topological insulators on fractal structures were studied \cite{yangPhotonicFloquetTopological2020, biesenthalFractalPhotonicTopological2022}, followed by acoustic higher-order topological states on fractals \cite{liHigherorderTopologicalPhase2022, zhengObservationFractalHigherorder2022, laiSpinChernInsulator2024}. The quantum spin Hall effect in a Sierpi\'nski gasket was observed in 2024, when spontaneously formed bismuth fractals on InSb were shown to have edge and corner states \cite{canyellasTopologicalEdgeCorner2024}.

Despite all these theoretical and experimental advances, the Haldane model on the Sierpi\'nski gasket has received less attention. In 2023, Li et al.~investigated such a model on a system consisting of two Sierpi\'nski gaskets glued together along one edge, and presented a squeezed version of the Haldane phase diagram \cite{liFractalityinducedTopologicalPhase2023}. 
The glued fractals form a more symmetric setup, which facilitates the analysis. However, the global self-similarity and the sublattice asymmetry are destroyed, thus modifying the gap structure.

In this work, we consider just one copy of the Sierpi\'nski gasket, keeping the asymmetry and the global self-similarity intact, and show that the phase diagram actually consists of more intricate and complex patterns. We implement a Haldane-like tight-binding model on two different geometries, a 2D triangular flake and the fractal Sierpi\'nski gasket, and show that the fractal system has two distinct types of band gaps, \textit{fractal gaps} and \textit{flux-induced gaps}. We investigate the behavior of both types of gaps and find that they can host topological modes for a wide range of parameters. These topological states are revealed by calculating a topological invariant. We have adopted a real-space Chern number, as the usual momentum-space Chern numbers are not applicable due to the finite size of these structures. Using this topological invariant, we find topological phase diagrams for the two gaps and observe complex patterns. In addition, we verify the resilience of these topological states to disorder. Our results show the richness of topology combined with fractality.

The outline of this paper is as follows. In Sec.~\ref{sec:model}, the Haldane model is introduced, and the two geometries, a triangular flake and the Sierpi\'nski gasket are discussed. Then, in Sec.~\ref{sec:spectral}, we present their energy spectra, examine the spectral properties, and investigate the localization of a few boundary states. In Sec.~\ref{sec:topology}, we characterize the topological nature of this model. To this end, we discuss the difficulties of the usual momentum-space Chern markers on the considered geometries and introduce the real-space Chern number. 
Thereafter, we study the emergence of the different topological gaps in terms of the parameters of the model, present the phase diagrams of this model, and compare them with earlier results. Finally, we probe the robustness of some topological states by including an arbitrary on-site disorder in Sec.~\ref{sec:disorder}.

\section{\label{sec:model}          The Fractal Haldane model}
The Haldane model is a tight-binding model on a honeycomb lattice with a nearest-neighbor (NN) hopping, a staggered (Semenoff) mass, and a complex NNN hopping. The Semenoff mass breaks the sublattice symmetry, while the complex NNN hopping breaks time reversal (TR) symmetry, even in the absence of a net external magnetic field \cite{haldaneModelQuantumHall1988}. From the tenfold way, we know that this combination of symmetries allows for topological phases in 2D \cite{ryuTopologicalInsulatorsSuperconductors2010}.
The corresponding Hamiltonian is
\begin{eqnarray}
    H &&=  t \sum_{\langle i, j \rangle}c^\dag_i c_j + M \sum_{i} \tau_i c^\dag_i c_i\nonumber\\
    &&+ \lambda \sum_{\langle \langle i, j \rangle \rangle}e^{-i\mu_{ij}\Phi}c^\dag_i c_j + \text{H.c.}.
    \label{eq:hamil}
\end{eqnarray} 
The first term describes the NN hopping of an electron with an amplitude $t$, the second term describes the staggered mass $M$ on the two sublattices, with $\tau_i = 1 (-1)$ when $i$ belongs to sublattice $A$ ($B$), and the last term describes a complex NNN hopping of strength $\lambda$. 
In this model, an electron acquires a complex phase $\mu_{ij}\Phi$ in a NNN hopping. Here, $\mu_{ij}$ is $1 (-1)$ for clockwise (counterclockwise) NNN hoppings and $\Phi$ is related to some intrinsic, periodic, and local flux density, which is constrained to be net-zero in a unit cell but breaks TR symmetry \cite{haldaneModelQuantumHall1988}. 

\begin{figure}
    \includegraphics[width=\linewidth]{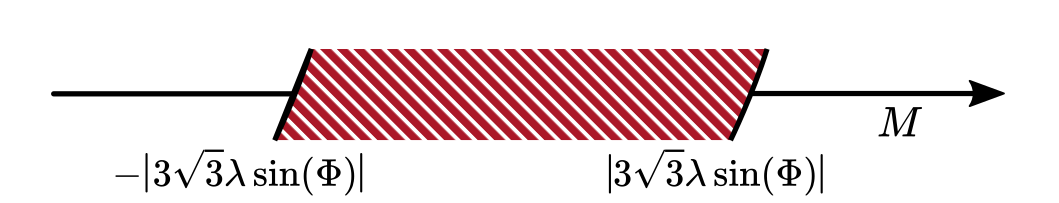}
    \caption{A topological phase diagram of the original Haldane model as a function of $M$. The red region corresponds to the conditions under which the system has topological modes.}
    \label{fig:phase}
\end{figure}

For $M=\lambda=0$, the system is gapless and semimetallic. The inclusion of a finite $\lambda$ ($M$) opens a topological (trivial) gap.
These effects compete and result in a topological phase transition when \cite{haldaneModelQuantumHall1988}
\begin{equation}
    M = \pm 3\sqrt3 \lambda \sin(\Phi).
    \label{eq:phase_boundary}
\end{equation}
As shown in Fig.~\ref{fig:phase}, when $\abs{M} < \abs{3\sqrt3 \lambda \sin(\Phi)}$, the system is a topological insulator, which is characterized by a nonzero quantized Hall conductance.
This means that if one introduces edges to the system, it will host gap-crossing modes, which are strongly localized on the edges. 
On the other hand, when $\abs{M} > \abs{3\sqrt3 \lambda \sin(\Phi)}$, the system is a normal insulator, with zero conductance everywhere. 
The transition between these phases goes through a conducting phase, in which the band gap is closed. This one-to-one relationship between the bulk properties and the behavior at the edges is known as the bulk-boundary correspondence, and is captured by the TKNN formula \cite{thoulessQuantizedHallConductance1982},
\begin{equation}
    \sigma_{H} = \frac{e^2}{h} \sum_n C^{(n)}.
\end{equation}
Here, $\sigma_H$ is the quantized Hall conductance at the edge, and $\sum_n C_n$ is the sum over the Chern numbers of all filled bands $n$. 
This Chern number is given by
\begin{equation}
    C^{(n)} = \frac{1}{2\pi} \iint_{BZ} F_{n}(\mathbf{k})d\mathbf{k}, 
    \label{eq:Chern}
\end{equation}
where $F_{n}(\mathbf{k})$ is the Berry curvature of the $n$th filled band, and the integral is over the first Brillouin zone \cite{hatsugaiChernNumberEdge1993, hasanColloquiumTopologicalInsulators2010}. 
Therefore, the Chern number is a bulk property, completing the bulk-boundary correspondence.

Haldane considered his model for the bulk of a honeycomb lattice, i.e.~with periodic boundary conditions (PBC). In this case, the system becomes translationally invariant and the momentum $\mathbf{k}$ is a good quantum number. Therefore, the Hamiltonian can be represented in $k$-space using Bloch's theorem \cite{blochUberQuantenmechanikElektronen1929}; the Chern number [Eq.~(\ref{eq:Chern})] is well-defined and integer-valued because the Brillouin zone forms a compact manifold.

In contrast, we investigate two lattices with open boundary conditions (OBC); a triangular flake and the fractal Sierpi\'nski gasket lattice. These systems are no longer translationally invariant, and therefore one cannot represent the Hamiltonian in $k$-space. Nevertheless, the bulk-boundary correspondence still holds, and one can predict the behavior at the edges if one has a well-defined bulk.

\subsection{\label{sec:fractal}     Geometries}
\begin{figure}
    \includegraphics[width=\linewidth]{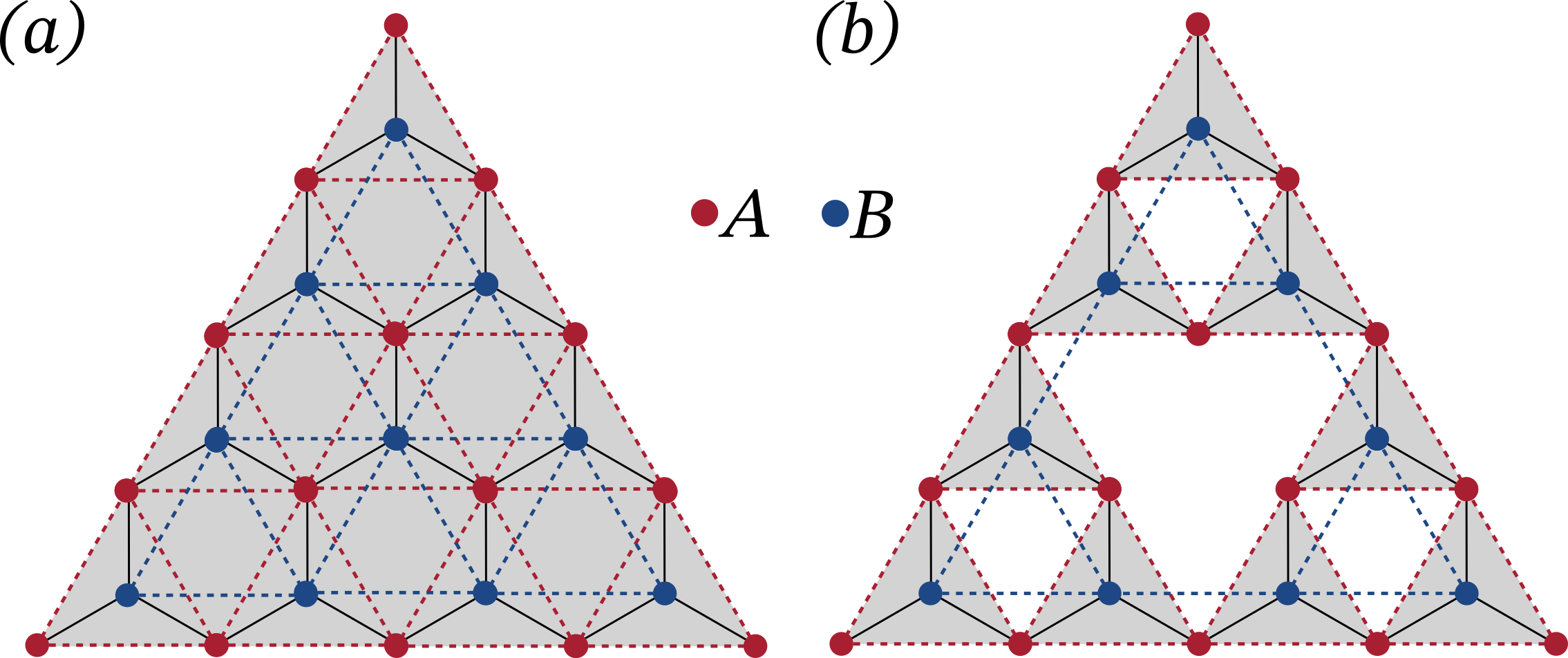}
    \caption{OBC lattice realizations corresponding to (a) a triangular flake and (b) a second-generation Sierpi\'nski gasket. Sublattices $A$ and $B$ are marked in red and blue, respectively. The NN hoppings are depicted as solid black lines and NNN hoppings are depicted as dashed lines with a color based on the sublattice that they connect.}
    \label{fig:FracLat}
\end{figure}

\textit{Triangular flake}. We have chosen a triangular flake [Fig.~\ref{fig:FracLat}(a)] to serve as a 2D 
counterpart to the Sierpi\'nski triangle [Fig.~\ref{fig:FracLat}(b)].
The two sublattices, $A$ and $B$ are represented in Fig.~\ref{fig:FracLat} in red and blue, respectively. Furthermore, the NN (solid) and NNN (dashed) hoppings are also depicted.

\textit{Sierpi\'nski Gasket}. The fractal that we consider in this paper is the Sierpi\'nski gasket. The Sierpi\'nski gasket is generated by subdividing an equilateral triangle into four congruent triangles and removing the central one (Fig.~\ref{fig:SG}). 
If one repeats this process $n$ times, one is left with the $n$th-generation Sierpi\'nski gasket. The Hausdorff dimension of this fractal is given by $d_H = \log{3}/\log{2} \approx 1.585$.
There are multiple ways to construct a fractal lattice model corresponding to the Sierpi\'nski gasket. 
In this work, we choose the method shown in Fig.~\ref{fig:FracLat}(b).
This construction method matches the triangular honeycomb lattice of Fig.~\ref{fig:FracLat}(a) when the sites in the voids of the $n$th generation of the Sierpi\'nski gasket are excluded. 

For both the triangular flake and the fractal lattice in Fig.~\ref{fig:FracLat}, the number of sites in each sublattice differs. This is a consequence of the choice of termination, which in this case consists of `zigzag' edges \cite{kondoQuantumSpinHall2019}. 

\subsubsection{NNN on a Sierpi\'nski gasket}
On the Sierpi\'nski gasket, the sets of NN and NNN hoppings are constrained by the geometry of the fractal. Specifically, the construction of the NNN hoppings is subject to a choice. 
As shown in Fig.~\ref{fig:FracLat}(b), all complex NNN hoppings between atoms in the $B$ sublattice are represented by lines crossing the voids of the fractal. 
Because these spaces do not exist in the fractal dimension, one could argue that these paths are restricted and these hoppings should not be considered.
We choose to include them since the Haldane model is often used to model SOC. An example would be the Kane-Mele model \cite{kaneQuantumSpinHall2005}, where each spin species is represented by one copy of the Haldane model. 
In this case, these complex NNN hoppings are viewed as NNN hoppings via the ``in-between site'', which is a second-order process. 
However, whether to include these void-crossing hoppings or not is highly dependent on the material and setup that one wants to model. Therefore, we present the results on a ``void-respecting'' Haldane model in Appendix \ref{app:voidHaldane}.

\section{\label{sec:spectral}        Spectral properties}
\subsection{\label{sec:energy}      Energy Spectra}
First, we investigate the energy spectra. As we are interested in the influence of fractality, the results on a Sierpi\'nski triangle are compared to those on a triangular flake. To keep the length scales consistent, a comparison is made between systems of equal side lengths.
Consequently, the triangular flake contains more lattice sites, as it has no holes. Explicit formulas for the number of sites of the Sierpi\'nski triangle $N_{fractal}$ and the triangular flake $N_{flake}$ of generation $n$ are given by
\begin{equation*}
    N_{fractal} = \frac{5}{2} 3^n + \frac{3}{2}, \qquad
    N_{flake} = (2^n + 1)^2 .
\end{equation*}
For example, a fourth-generation fractal contains 204 sites, and a triangular flake of a similar size contains 289 sites. This difference grows quickly with $n$.

\begin{figure}
    \includegraphics{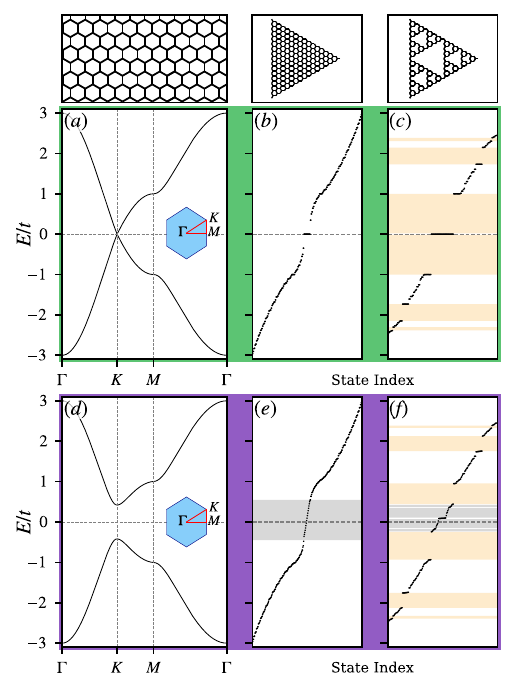}
    \caption{The energy spectra of (a), (d) the bulk Haldane model, (b), (e) a triangular flake, and (c), (f) a fourth-generation Sierpi\'nski gasket for both the semimetallic phase (green), $\lambda = M = 0t$, and a topological bulk phase (purple), $\lambda = M = 0.1t$. For the bulk system, the path through $k$-space is shown in an inset. Here, we observe the characteristic Dirac cone of graphene in (a), which becomes gapped for a topological phase, depicted in (d). For the flake, we see a continuous spectrum in both the semimetallic (b) and topological (e) phases, however in the case of a topological phase some states lie in the bulk gap. 
    The fractal case is more interesting because the system becomes gapped even in the semimetallic phase (c); we will refer to these gaps as the \textit{fractal gaps}. These gaps are marked in orange.
    When $\lambda = M = 0.1t$ (f), we see that most degeneracies of (c) are broken, except for the middle band. This degeneracy is only decreased as some states move away from the flat band at $E=0.1t$.
    Both the fractal and the triangular flake show a flat band in the semimetallic phase, as a consequence of their bipartite nature and the difference in the number of sites in each sublattice.}
    \label{fig:Spectra}
\end{figure}

In Fig.~\ref{fig:Spectra}, the energy spectrum of (a) the bulk Haldane model, (b) a triangular flake, and (c) a fourth-generation Sierpi\'nski triangle are shown for the semi-metallic phase with $\lambda = M = 0$. Similarly, in Figs.~\ref{fig:Spectra}(d)-(f), the same geometries are considered for a topological phase with $\lambda = M = 0.1t$. The influence of the Semenoff mass and the complex NNN hopping are also investigated individually and these results are presented in Appendix \ref{app:individualTerms}.

In Fig.~\ref{fig:Spectra}(a), the characteristic Dirac cone around the $K$-point of the bulk is depicted.  In Fig.~\ref{fig:Spectra}(b), the spectrum of a triangular flake is shown. At first glance, it might appear to be gapped because there seem to be no states around $E=0$. However, calculations performed for generations $g = 3,4,5$ indicate that this is a finite-size effect and the gap vanishes in the thermodynamic limit \cite{akhmerov_boundary_2008}, see Appendix \ref{app:finiteSize}.
In the fractal system [Fig.~\ref{fig:Spectra}(c)], we observe new spectral gaps between $E = 0$ and $E = \pm 1t$, and around $E=\pm 2t$. These gaps are neither present in the spectrum of the bulk nor of the triangular flake. They are a consequence of fractality and remain robust in the thermodynamic limit (see Appendix \ref{app:finiteSize}). We will refer to them as \textit{fractal gaps}. These fractal gaps are marked in orange.
Furthermore, we notice a central flat band in both OBC cases [Figs.~\ref{fig:Spectra}(b) and (c)], which is caused by the bipartite nature of these lattices in the absence of a complex NNN hopping \cite{liebTwoTheoremsHubbard1989, conteFractalLatticeHubbardModel2023, liFractalityinducedTopologicalPhase2023}. Additionally, in the case of the Sierpi\'nski gasket, multiple flat bands are observed away from $E=0$. These are comprised of compact localized states, as discussed in Ref.~\cite{conteFractalLatticeHubbardModel2023}. 
In Figs.~\ref{fig:Spectra}(d)-(f), analogous plots are made, but now including a complex NNN hopping and a staggered mass $\lambda = M =0.1t$, with complex NNN hopping phase $\Phi = \pi/2$. 
Under these conditions, the bulk becomes gapped, as shown in Fig.~\ref{fig:Spectra}(d). Upon consideration of the OBC systems, three notable features become apparent.

\paragraph{Broken Symmetry.} First, we see that the spectra of the triangular flake and the Sierpi\'nski gasket [Figs.~\ref{fig:Spectra}(e) and \ref{fig:Spectra}(f), respectively] are slightly asymmetric around $E=0$ (dashed gray horizontal lines). 
This is a consequence of a nonzero staggered mass and the unbalanced number of lattice sites in each sublattice \cite{liebTwoTheoremsHubbard1989}. 
However, there is a fundamental distinction to be made between the fractal and the triangular flake. 
In the thermodynamic limit of the latter, the ratio of the number of sites in each sublattice tends to one, resulting in a negligible difference. This is because of the difference in the number of sites scales with the total edge length, while the total number of sites scales with the area. 
On the other hand, for the Sierpi\'nski gasket, the ratio between the number of sites in each sublattice tends to $1.5$, making this difference and the corresponding asymmetric energy spectrum not just an outer-edge effect, but an intrinsic property of this Sierpi\'nski gasket construction.

\paragraph{Flat bands.}
Second, we observe that the central flat band is lost when comparing the spectra of the triangular flake without and with the mass and the complex NNN hopping [Figs.~\ref{fig:Spectra}(b) and \ref{fig:Spectra}(e)]. This loss of degeneracy occurs because the introduction of (complex) NNN hoppings breaks the bipartite nature of this lattice. However, when a similar comparison is made between Figs.~\ref{fig:Spectra}(c) and \ref{fig:Spectra}(f) for the fractal, the central flat band remains, even though the bipartite nature of the lattice is broken. 
All other flat bands depicted in Fig.~\ref{fig:Spectra}(c) are broken in Fig.~\ref{fig:Spectra}(f).

\paragraph{In-gap states.} Lastly, one more observed effect is the emergence of states within the different spectral gaps of these systems. For the triangular flake, we observe a set of in-gap states crossing the bandgap of the bulk model of Fig.~\ref{fig:Spectra}(e). 
In the case of the Sierpi\'nski gasket, a more nuanced process takes place. The introduction of a complex NNN hopping causes the central, originally flat band to partially disperse. This transforms it into a set of states which have moved into the largest fractal gaps, without closing these gaps. This contrasts with the triangular flake, where the entire gap is populated with topological modes.

Although the fractal gaps are never closed and remain trivial for $\lambda=M=0.1t$, this does not mean that the fractal does not host topological states. 
Previous investigation of similar fractal systems showed that these new non-degenerate states can have a topological character \cite{yangPhotonicFloquetTopological2020, biesenthalFractalPhotonicTopological2022}. Some of the dispersed modes of the central band can form new ``bulk bands'' away from $E=0$. Therefore, the introduction of a Haldane flux will cause the emergence of new gaps in the spectrum of this fractal, which we will call \textit{flux-induced gaps} (marked in gray). These flux-induced gaps could in principle host topological ``in-gap'' states and the spreading of the states in the energy spectrum shown in Fig.~\ref{fig:Spectra}(f) suggests that this is the case.

The difference in behavior between the triangular flake and the fractal system is striking. The triangular flake was not gapped for $\lambda=M=0$, but has a finite bulk gap populated by topological edge modes when $\lambda = M = 0.1t$. On the contrary, the Sierpi\'nski gasket had (multiple) flat band(s) and multiple fractal gaps for $\lambda = M =0$. For the parameters in Fig.~\ref{fig:Spectra}(f), the fractal gaps have not been closed, but instead, the flat band has split and the resulting flux-induced gaps are populated by ``in-gap'' states. Only upon further increasing $\lambda$ the fractal gap will close and reopen too, now populated by ``fractal gap''-crossing edge states.

\subsection{\label{sec:TopoStates} Edge States}
\begin{figure}
    \includegraphics{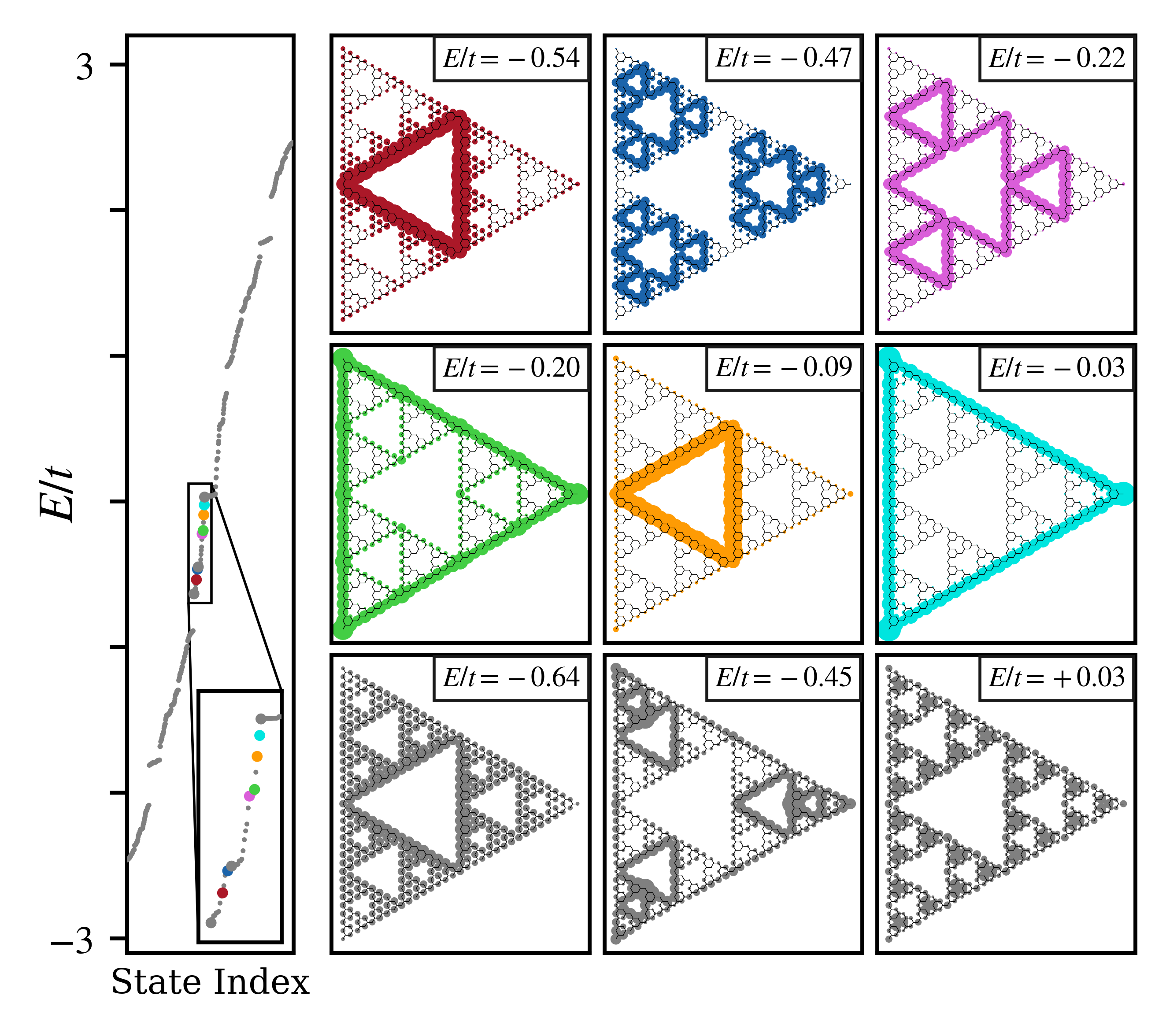}
    \caption{The probability density $|\psi|^2$ of some states on a fifth-generation Sierpi\'nski gasket, with a nonzero complex NNN hopping. Here, $\lambda = 0.2t$ and $M = 0.05t$. These states correspond to points lying in between the bands depicted in Fig.~\ref{fig:Spectra}(f). Each state's energy is depicted as a point of the corresponding color in the energy spectrum. The colored states show strong localization on the different edges of the Sierpi\'nski gasket. The gray states lack this edge localization and are ``bulk'' states.}
    \label{fig:TopoStates}
\end{figure}

To corroborate our claims in the previous section, we briefly investigate the spatial profile of the states in the flux-induced gaps. In Fig.~\ref{fig:TopoStates}, a subset of these states is shown, distinguished by their strong localization on the different edges of a fifth-generation Sierpi\'nski gasket (colored) or lack thereof (gray).
As a consequence of the many voids of the Sierpi\'nski gasket, the lattice contains more edges. 
Subsequently, edge modes (colored) are observed to localize on different sets of edges, ranging from edges around the smallest voids to the outermost edges, and combinations of these.

\section{\label{sec:topology} Topological Characterization}
In this section, we will introduce the topological invariant that will be used to characterize the topology of the different (fractal or flux-induced) gaps of the Sierpi\'nski gasket. Usual calculation methods of topological invariants assume translational invariance, which is not present in a fractal. Moreover, the notion of a bulk-boundary correspondence becomes ill-defined, as a Sierpi\'nski gasket has no bulk, due to voids on each length scale. Therefore, we will consider the local real-space method proposed by Kitaev to calculate a topological invariant \cite{kitaevAnyonsExactlySolved2006}.
Previous studies have applied this invariant to characterize the quantum Hall effect in fractals. Here, good agreement between this invariant and the transverse Hall current (obtained by direct transport calculations) is observed \cite{fremlingExistenceRobustEdge2020, fischerRobustnessChiralEdge2021}.

\subsection{\label{sec:RSCN}        Real-Space Chern number}
\begin{figure*}
    \includegraphics[width=\textwidth]{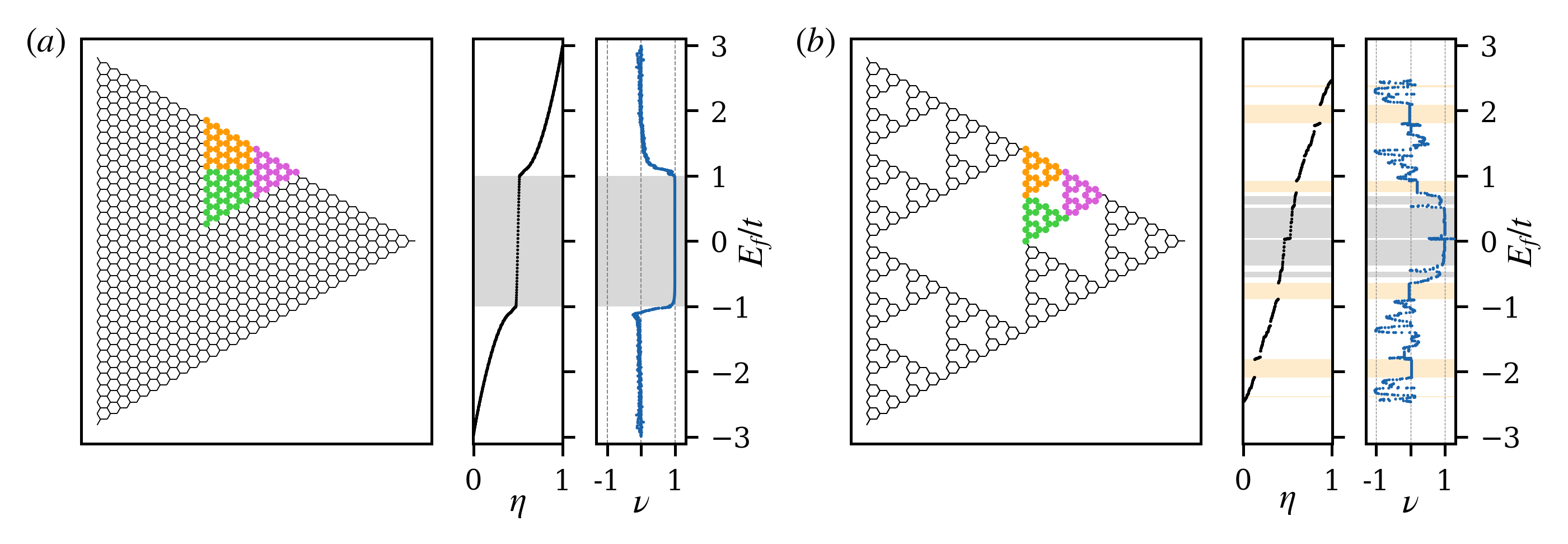}
    \caption{Real-space Chern number of $(a)$ a triangular flake and $(b)$ a Sierpi\'nski gasket. For both geometries, we set the complex NNN hopping strength $\lambda = 0.2t$, the complex phase factor $\Phi = \pi/2$ and the Semenoff mass $M = 0.05t$. We have depicted the choice of $A, B,$ and $C$ in pink, orange, and green, respectively. Furthermore, the energy spectra are shown as a function of the filling fraction $\eta$, and the RSCN $\nu$ is shown as a function of the Fermi level.}
    \label{fig:RSCN}
\end{figure*} 
The local real-space Chern number (RSCN), proposed by Kitaev, is given by \cite{kitaevAnyonsExactlySolved2006} 
\begin{equation}
    \nu({P}) = 12\pi i \sum_{j \in A}\sum_{k \in B}\sum_{l \in C} \left(P_{jk}P_{kl}P_{lj} - P_{jl}P_{lk}P_{kj}\right).
    \label{eq:RSCN_Kitaev}
\end{equation}
Here, $P$ is the projector onto the Fermi sea, and $A$, $B$, and $C$ are three touching subsets of lattice points arranged in counterclockwise order. 
Therefore, $P_{jk}$ is the ${j, k}$th element of the projector $P$ onto the occupied states at a given Fermi energy, when $P$ is written in the basis of local atomic orbitals.
When the bulk Chern number of the investigated system is quantized, $\nu(P)$ becomes independent of the specific choice of $A$, $B$, and $C$ in the limit where the number of sites in each subset approaches infinity. Therefore, $A$, $B$, and $C$ should be chosen such that each region is large enough to capture the system's topology. However, they cannot span the entire system because in that case $\nu({P}) = 0$, per construction \cite{biancoMappingTopologicalOrder2011}. 

In Fig.~\ref{fig:RSCN}, the choice of $A$ (pink), $B$ (orange), and $C$ (green), the energy spectrum, and the corresponding values of $\nu(P)$ are depicted for a triangular flake [Fig.~\ref{fig:RSCN}(a)], and for a fifth-generation Sierpi\'nski gasket [Fig.~\ref{fig:RSCN}(b)], with $\lambda=0.2t$ and $M = 0.05t$. The energy spectra are shown in terms of the Fermi energy and the corresponding percentage of filled states $\eta$, which, from now on, will be referred to as the ``filling fraction''. 

In Fig.~\ref{fig:RSCN}(a), we observe that the RSCN $\nu(P)$ is a quantized integer for $|E_f|<t$ and sharply drops to zero for Fermi energies residing in the bulk bands, where $|E_f|>t$. 
In Fig.~\ref{fig:RSCN}(b), multiple features are worth noting. First, the fractal gaps remain trivial [$\nu(P) = 0$] under these conditions. Second, there is a non-trivial region [$\nu(P) = 1$] around $E = 0$, as well as smaller topological regions [$\nu(P) = -1$] around other energies, $|E_f| \approx 1.35t $ and $|E_f| \approx 2.3t $. 
Upon closer inspection, the topological regions around $E_f = 0$ seem to coincide with the spreading of states in the flux-induced gaps. 
Another interesting feature is the sharp drop of the RSCN at $|E_f| = 0.6t$, which tells us that not all of the states that moved out of the flat band are topological, but some form a new ``bulk band'' at an energy away from $E=0$. 
We also note that at $E_f = M = 0.05t$, the spectrum forms a flat band and the value of $\nu(P)$ diverges, but this is instantly resolved outside the flat band. Because these RSCNs were calculated for the same parameters as considered in Fig.~\ref{fig:TopoStates}, they corroborate the depicted states as topological edge modes.
\begin{figure}
    \includegraphics[]{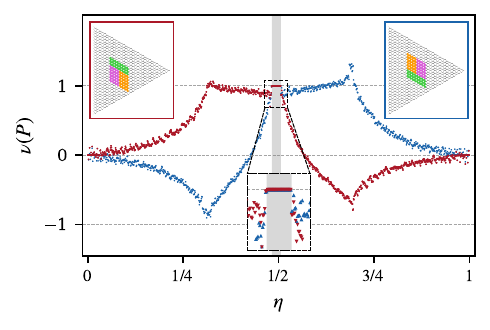}
    \caption{The numerically calculated values of $\nu(P)$ as a function of a projector onto the states filling the lowest $\eta$ percentile of states. The red and blue values correspond to the two choices of $A$, $B$ and $C$, depicted in the insets with the respective border color. In both cases, the same system was considered with $M=0.4t$, $\lambda=0.2$, and $\Phi=\pi/2$. The observed difference is a consequence of an undefined Chern number when the Fermi energy lies within a band.}
    \label{fig:RSCN_berryCurvature}
\end{figure}

Another curious artefact is the small bump in Fig.~\ref{fig:RSCN}(a), which can be observed at energies around $E_f=-t$. Further investigation of this bump reveals that this behavior is related to the specific choice of $A$, $B$, and $C$. In Fig.~\ref{fig:RSCN_berryCurvature}, $\nu(P)$ is shown as a function of the projector onto all states below the Fermi energy, which corresponds to a system with filling fraction $\eta$, for two different choices of $A$, $B$, and $C$. Here, the strength of the staggered mass is also increased to $M=0.4t$ to inflate the effect.
\begin{figure*}
    \includegraphics{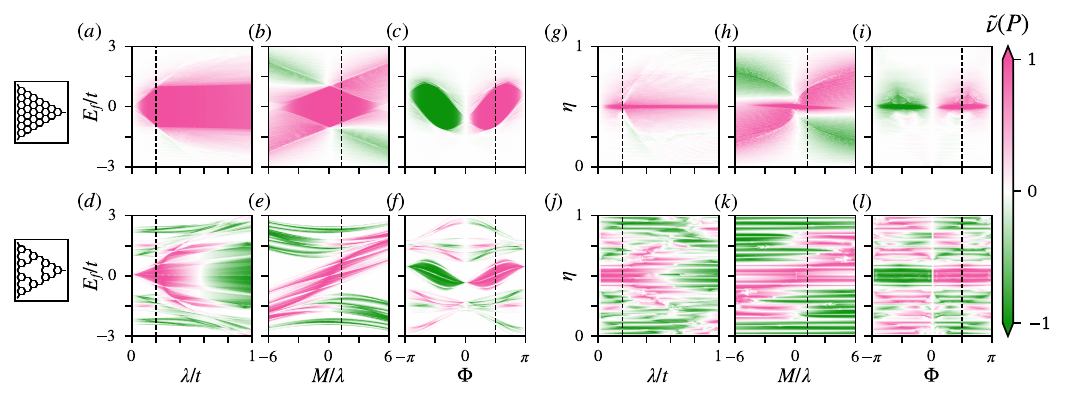}
    \caption{The interplay between $\lambda$, $M$ and $\Phi$ and the energy of the observed topological band gaps for a triangular flake [(a)-(c) and (g)-(i)] and a Sierpi\'nski gasket [(d)-(f) and (j)-(l)]. In (a)-(f), the energy of a gap is considered, while the filling fractions are considered in (g)-(l). For each figure, one parameter is varied, while the others are set to $\lambda =0.2t$, $M=0.05t$, and $\Phi =\pi/2$. These values are marked with a black dashed line in the respective figures. Both systems have $t=1$, and have the same size. A deep pink (green) color represents a stable value of $\tilde{\nu}(P) = 1$ $(-1)$. In (a)-(c) and (g)-(i), the opening and closing of a topological gap under the conditions predicted by Haldane \cite{haldaneModelQuantumHall1988} can be observed. In (d)-(f) and (j)-(l), we observe the opening and closing of multiple topological gaps, also away from $E_f=0$ and $\eta = 0.5$. These topological gaps have a much more complex dependence on the parameters. }
    \label{fig:Butterflies}
\end{figure*}
The two choices of $A$, $B$, and $C$ are depicted in the insets and the corresponding values of $\nu(P)$ are shown in red and blue, respectively. 
For most filling fractions, the red and blue graphs do not coincide, except for a small range of filling fractions around $\eta = 1/2$ (marked in gray). This range corresponds to the topological states, which have an RSCN quantized exactly at $\nu(P) = 1$. The RSCN is independent of the specific choice of $A$, $B$, and $C$ when the Chern number [Eq.~(\ref{eq:Chern})] itself is quantized. A quantized Chern number exists only as a property of a full band, but for filling fractions inside the bulk bands (outside the gray area), the Chern number is not defined and consequently $\nu(P)$ is non-physical.

In a sense, Fig.~\ref{fig:RSCN_berryCurvature} shows the real-space equivalent of the process of summing over the Berry curvature. How and which different choices of $A$, $B$, and $C$ compose different routes to the same topological invariant is an interesting topic of research which requires more investigation. For now, a method to determine whether a filling fraction constitutes a system with filled ``bulk bands'' is needed.
Here, we offer two possible solutions. The first method is to consider a large set of choices for $A$, $B$, and $C$, and verify whether the values of $\nu(P)$ are in agreement, but this can be costly.
A second approach is to take a numerical derivative with respect to $\eta$ and check whether the rate at which the value of $\nu(P)$ changes is small. A change of $\nu(P)$ implies that it is not yet quantized, the mode is part of a band, and the Chern number is ill-defined. However, this method might lose some ``in-gap'' modes close to the boundaries of topological gaps, due to the nature of a derivative. Nevertheless, since this approach requires significantly less computational time, we adopt it here. Hence, we introduce $\tilde{\nu}(P) $, which is a masked version of $\nu(P)$. This mask exponentially fades out values of the RSCN with a large numerical derivative of $\nu(P)$, with a half-life of $\nu'_{1/2} = 0.05$. Its effect is discussed in Appendix~\ref{app:visualMask} in more detail.

\subsection{\label{sec:butterfly}      Sierpi\'nski cocoons}
Now, we will investigate the topological gaps and the conditions under which they arise. We consider a fifth-generation Sierpi\'nski gasket and an equally sized triangular flake and set $t=1$, such that we are left with $\lambda$, $M$ and $\Phi$. The RSCN of the systems as a function of $\lambda$, $M$, or $\Phi$ and the Fermi energy $E_f$ is shown in Figs.~\ref{fig:Butterflies}(a)-(f), for both the triangular flake and the Sierpi\'nski gasket. The plots are reminiscent of Hofstadter butterflies, obtained when considering the quantum Hall effect (see Refs.~\cite{hofstadterEnergyLevelsWave1976, kooiGenesisFloquetHofstadter2018}). Because of this similarity, we will refer to these as Sierpi\'nski cocoons, which are primordial to the Hofstadter butterfly. While the Hofstadter butterflies arise when plotting the energy gaps as a function of an external magnetic flux, here, the parameters of the model are of $t$, $\lambda$, $M$, and $\Phi$.
In each Sierpi\'nski cocoon, either $\lambda$, $M$, or $\Phi$ are varied, while the other parameters are chosen to be $\lambda = 0.2t$, $M=0.05t$ and $\Phi = \pi/2$. These values are marked with a black dashed line in the respective figures. 
In Fig.~\ref{fig:Butterflies}(a), we see that only for $\lambda > M/3\sqrt{3}$ there exists a well-defined topological gap in the spectrum of a triangular flake, in agreement with the bulk Chern number of the Haldane model. 
In Fig.~\ref{fig:Butterflies}(b), it is shown that the topological gap is closed for a $\abs{M} > 3\sqrt{3}\lambda$, and in (c), these topological gaps show a sinusoidal dependence on $\Phi$, as expected from the results by Haldane \cite{haldaneModelQuantumHall1988}. 
There are also some differences with the Haldane model: in Figs.~\ref{fig:Butterflies}(a)-(c), a lack of symmetry around $E=0$ is of note, which is caused by a nonzero staggered mass, in combination with the asymmetric number of sites in each sublattice. Additionally, the numerical nature of the RSCN calculation is apparent in these figures in the form of areas of numerical noise.  
Furthermore, the four (diagonal) white lines in Fig.~\ref{fig:Butterflies}(b) are caused by the masking of $\nu(P)$, and they correspond to the diverging bump in $\nu(P)$, as discussed in Sec.~\ref{sec:RSCN}.

The results of a similar procedure on a fifth-generation Sierpi\'nski triangle are depicted in Figs.~\ref{fig:Butterflies}(d)-(f). A much richer structure is observed. 
When comparing the influence of complex NNN hopping strength on the topological bands in Figs.~\ref{fig:Butterflies}(a) and (d), more distinct topological gaps are observed in the case of a Sierpi\'nski triangle. 
For any small value of $\lambda/t$, the model exhibits a central band gap with $\tilde{\nu}(P) = +1$. This is different from the triangular flakes, for which a minimal complex NNN hopping strength $\lambda = M/3\sqrt{3}$ was required to open a topological gap when $M\neq0$.
For larger values of $ 0 < \lambda/t < 0.6$, it becomes apparent that the central gap is formed of multiple smaller topological gaps, all with $\tilde{\nu}(P) = +1$. These gaps seem to close and reopen with the opposite Chern number for $0.6 < \lambda/t < 1$. This is in stark contrast with the results on a triangular flake, where just a single topological gap opens around $E_f = 0$, and no phase transition is observed.
Furthermore, there are also topological band gaps away from the Fermi level, with different Chern numbers.
The closing and reopening of topological gaps is a recurrent pattern, also visible in the topological bands around $E \approx 1.35t$ for smaller values of $\lambda/t$. However, in this case the topological gaps change from $\tilde{\nu}(P)  = -1$ to $ \tilde{\nu}(P) = 1$. 
In Fig.~\ref{fig:Butterflies}(e), the Semenoff mass is shown to impact the width of these central band gaps and to lead to a slanted, diagonal band.  The latter is a consequence of the imbalance in the number of sites in each sublattice and the different effect that the mass has on each sublattice. Again, topological gaps appear at energies away from $E_f = 0$. The structure acquires the shape of a letter `$Z$', pierced by multiple gaps.
Lastly, in Fig.~\ref{fig:Butterflies}(f) the more dense cocoon shape from Fig.~\ref{fig:Butterflies}(c) becomes smaller and consists of multiple distinct gaps. Once again, we observe more bands and topological gaps of different RSCN at higher energies. 

In Figs.~\ref{fig:Butterflies}(g)-(l), similar figures are displayed as in Figs.~\ref{fig:Butterflies}(a)-(f), but now the filling fraction $\eta$ is considered instead of the Fermi energy. 
They are related by the density of states, and therefore share many similarities. 
When compared to Figs.~\ref{fig:Butterflies}(a)-(f), we observe an important difference; the energy around which the topological gaps lie fluctuates, depending on the specific choice of parameters, but the filling fractions corresponding to topological gaps seem to be more stable, forming horizontal lines. In the case of a triangular flake, the topological gaps are found around half-filling, $\eta=0.5$, as expected. A similar effect is observed for the Sierpi\'nski gasket, most prominently in Fig.~\ref{fig:Butterflies}(k), where a dependence of the energy of a gap on the parameter $M$ is converted into a constant value of $\eta$. 
For the Sierpi\'nski gasket, there is a flat band around half-filling, and these states are non-topological, as best observed in Fig.~\ref{fig:Butterflies}(k). However, for any nonzero $\lambda$, the flux-induced gaps on either side of half-filling are topological, and these gaps are well-defined for all parameters.

There is one more distinction that we want to make to set the fractal results apart from the ones on either the triangular flake or the original bulk model.
For the fractal, as long as $\lambda > 0$, all parameters $\lambda$, $M$, or $\Phi$ can independently drive the different gaps of these structures through all phases. By only changing one parameter, the system can be tuned to go from a topological phase with $\tilde{\nu}(P) = 1$ to a trivial phase and then even to a phase with $\tilde{\nu}(P) = -1$. This is in contrast with the triangular flake and bulk model, where $\Phi$ determined the sign of the Chern number and $\lambda$ and $M$ only determined whether the system was in a topological phase or not.

\subsection{\label{sec:phase}       Phase diagram}
\begin{figure}
    \includegraphics{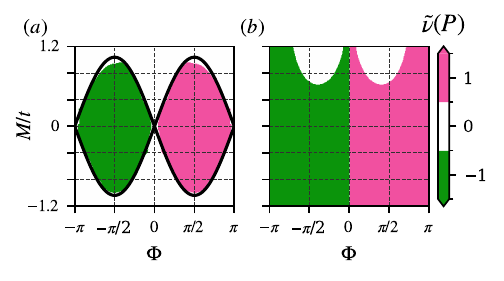}
    \caption{The phase diagram of (a) a triangular flake and (b) a fifth-generation Sierpi\'nski gasket for a constant filling fraction $\eta$, with complex NNN hopping strength $\lambda=0.2t$, as a function of the staggered mass $M$ and the complex NNN hopping phase $\Phi$. For the triangle, $\eta=0.5$ and for the Sierpi\'nski gasket 
    $\eta \approx 0.46$. The colors represent the rounded RSCN. The theoretical phase boundaries given in Eq.~(\ref{eq:phase_boundary}) are represented in black, showing good agreement with the triangular model considered, up to some finite-size effects.}
    \label{fig:phase_diagram_comparison}
\end{figure}

To compare the results of this work with the ones by Haldane~\cite{haldaneModelQuantumHall1988} and by Li et al.~\cite{liFractalityinducedTopologicalPhase2023}, we present a phase diagram as a function of the staggered mass $M$ and the complex NNN hopping phase factor $\Phi$. 
Since the gaps are better defined in terms of a constant filling fraction $\eta$ than in terms of an energy $E$, as depicted in Figs.~\ref{fig:Butterflies}, the phase diagrams are shown for systems of constant $\eta$. For the triangular flake, we choose $\eta = 0.5$ because the topological gap is predicted to be at half filling and Figs.~\ref{fig:Butterflies}(g)-(i) show a well-defined topological gap for this filling, for a wide range of both $M$ and $\Phi$. However, for the Sierpi\'nski gasket we select $\eta = 281/609 \approx 0.46$. This is less than half-filling to circumvent the observed flat band. Moreover, from Figs.~\ref{fig:Butterflies}(j)-(l) we know that there is a well-defined topological band gap for this fraction for a wide range of $M$ and $\Phi$. 
We have chosen these values of $\eta$ because both of them correspond to a gap opened by a finite complex NNN hopping. A more direct comparison with the same values of $\eta$ would be desirable but the fractal nature of the Sierpi\'nski gasket makes such a comparison impossible because of the emergence of a central flat band consisting of compact localized states \cite{conteFractalLatticeHubbardModel2023}.
In Fig.~\ref{fig:phase_diagram_comparison}, the phase diagrams are shown for (a) a triangular flake and (b) a Sierpi\'nski gasket of the same size.

\begin{figure}
    \includegraphics[width=0.7\linewidth]{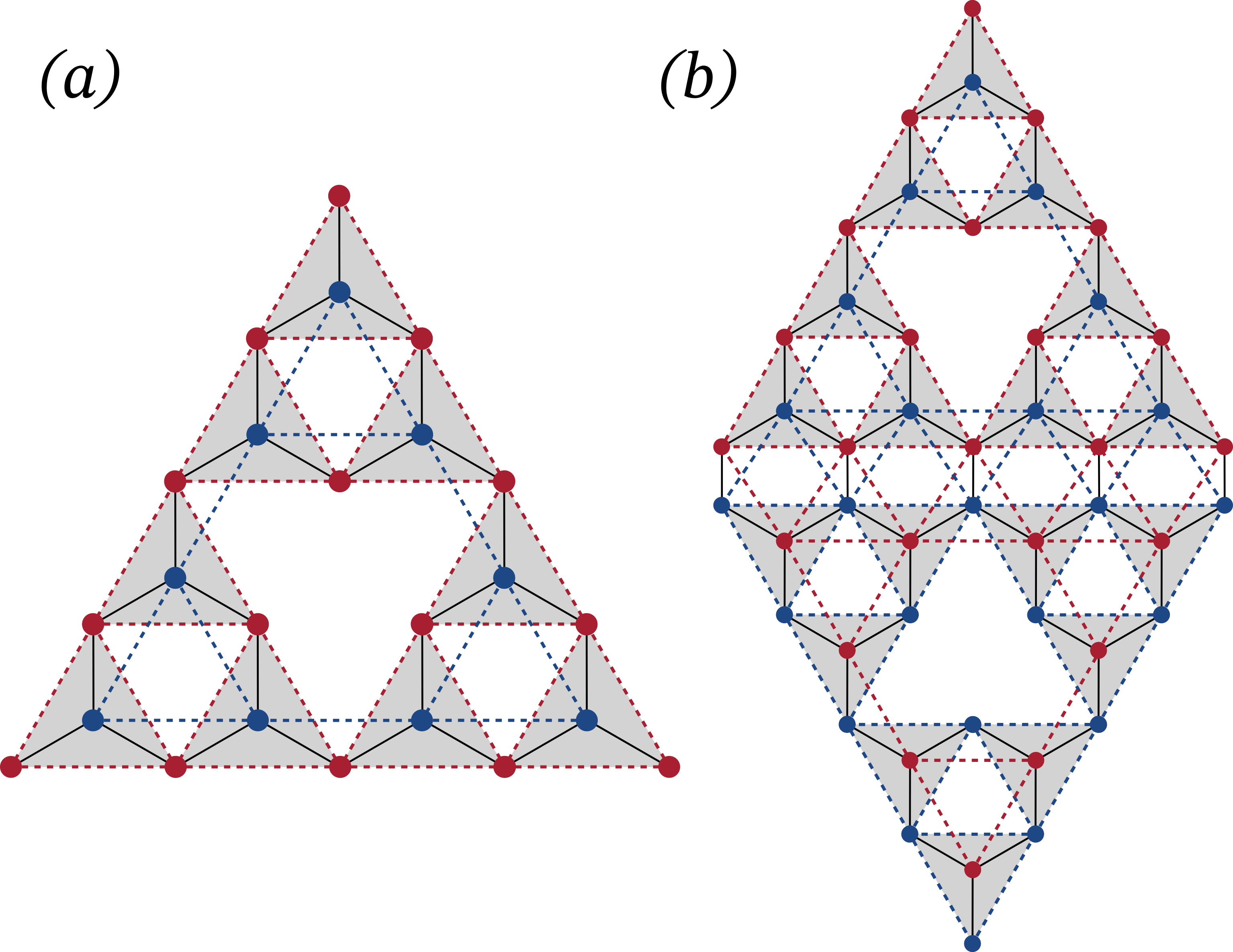}
    \caption{A second generation of (a) the single Sierpi\'nski gasket considered in this study and (b) the doubled fractal investigated in Ref.~\cite{liFractalityinducedTopologicalPhase2023}. The doubled Sierpi\'nski gasket restores the sublattice symmetry and the global self-similarity is lost.}
    \label{fig:comparison}
\end{figure}

The expected phase boundaries, as predicted by Haldane \cite{haldaneModelQuantumHall1988}, are depicted by the black lines. These predicted boundaries are a good fit for the triangular lattice in Fig.~\ref{fig:phase_diagram_comparison}(a). However, for the Sierpi\'nski gasket, we see that the phase diagram at $\eta \approx 0.46$ does not fit these phase boundaries
at all. This is in stark contrast with the earlier work by Li et al., who presented a squeezed Haldane phase diagram for two glued Sierpi\'nski gaskets \cite{liFractalityinducedTopologicalPhase2023}. It is important to keep in mind the differences between the model studied here and the one in Ref.~\cite{liFractalityinducedTopologicalPhase2023}: we consider a single Sierpi\'nski gasket, while two Sierpi\'nski gaskets connected along an entire edge were considered in Ref.~\cite{liFractalityinducedTopologicalPhase2023}. In Fig.~\ref{fig:comparison}, we show both structures: a single and a doubled Sierpi\'nski gasket, in (a) and (b), respectively.

\begin{figure*}
    \includegraphics{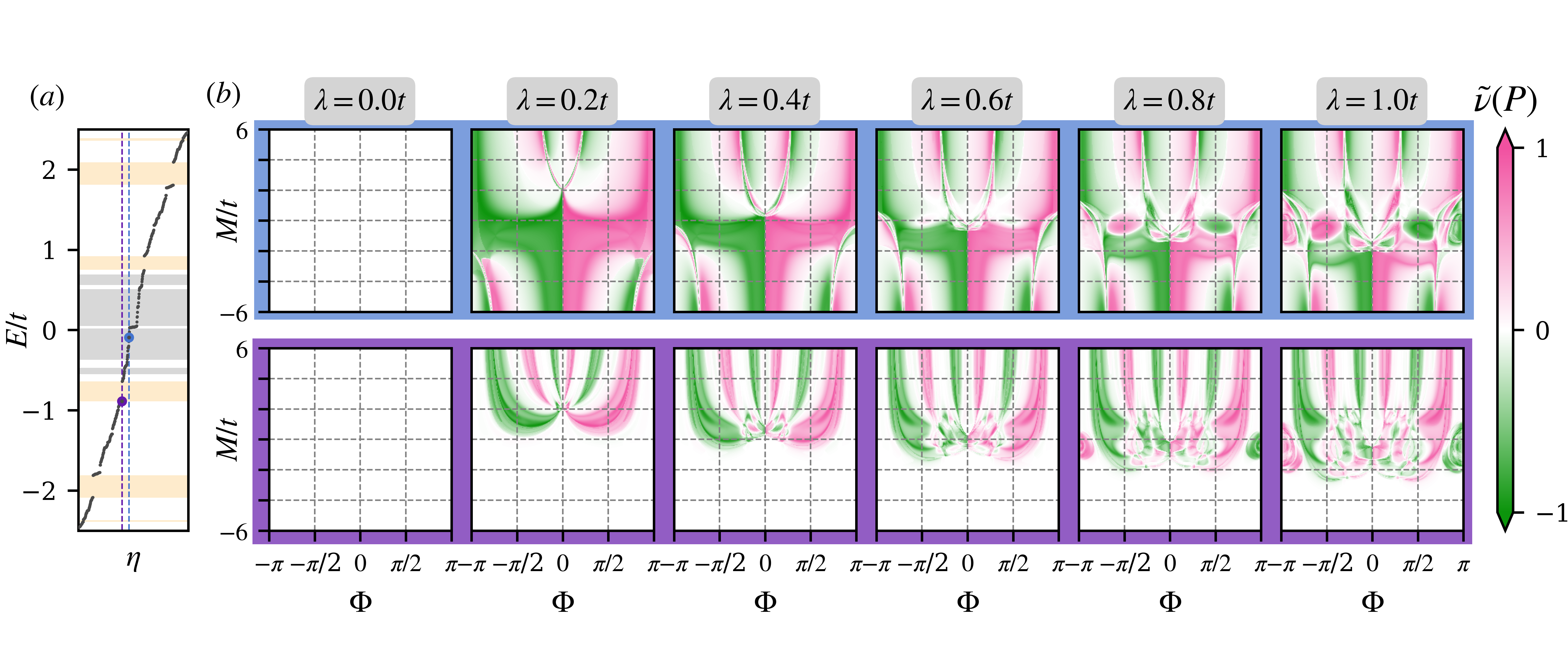}
    \caption{The phase diagrams corresponding to two different filling fractions of a fifth-generation Sierpi\'nski triangle. These two are $\eta=281/609$ and $\eta = 242/609$, corresponding to a flux-induced and a fractal gap, respectively. In (a), the energy spectrum is shown for $M=0.01t$ and $\lambda=0.12t$, and the two filling fractions $\eta$ of interest are marked with blue and purple lines, together with the last filled state. In (b), the RSCNs of these gaps are shown for a set of complex NNN hopping strengths ranging from $0$ to $t$. The masked RSCN $\tilde{\nu}(P)$ is represented as a color map ranging from 1 to -1. For both filling fractions, we see the emergence of topological phases. In the top row, topological phases are observed for a wide range of parameters, showcasing intricate patterns. For the largest fractal gap, we also observe topological phases, with a similar complexity. This interesting behavior increases with the strength of NNN hoppings and is reminiscent of Arnold's tongues. }
    \label{fig:Fractal_Phase}
\end{figure*}
If one considers such a doubled fractal, as in Ref.~\cite{liFractalityinducedTopologicalPhase2023}, two features are lost. First, the symmetry between the two sublattices is restored. This is because the roles of the $A$ and $B$ sublattices are reversed in the mirrored copy of the Sierpi\'nski gasket, such that the total number of sites in each sublattice is equal.
Second, although the doubled fractal indeed has a fractal dimension of $d_H \approx 1.58$, it has lost the global self-similarity of the Sierpi\'nski fractal. 

Furthermore, the authors of Ref.~\cite{liFractalityinducedTopologicalPhase2023} (and many others) quantify the topology of a fractal with the Bott index. 
The Bott index is a \textit{global} quantity describing the bulk.
In general, to apply the Bott index one implements (artificial) PBC, and one then performs a \textit{global} calculation, after which the behavior at the edge is inferred via the bulk-boundary correspondence.
However, compared to more traditional OBC structures, the Sierpi\'nski gasket has no well-defined bulk because it contains not only an outermost edge, but also many internal edges on the boundaries of the voids. 
Finding boundary conditions that eliminate these internal edges is quite complicated. If this is not done, these holes and their edges remain in the Bott index calculation.
Furthermore, applying periodic boundary conditions to a Sierpinksi gasket introduces a new length scale, which breaks the self-similarity and the dimensionality on scales larger than this length scale. This change of the dimensionality could have non-trivial effects.
Taken together, it is questionable whether this marker indeed describes the behavior on the internal edges or at the boundaries of a larger lattice of these fractals. 
On the other hand, the RSCN, provides \textit{global} information based on a \textit{local} calculation, which does not require the notion of a well-separated bulk. Consequently, the RSCN method only requires the \textit{local} wavefunctions in real space for the calculation of the topological invariant, and is more adequate than the Bott index in the case of fractals.

As shown in Fig.~\ref{fig:Spectra}, the fractality of the Sierpi\'nski gasket opens trivial gaps in the spectrum, and a complex NNN hopping can push the system into a topological phase. In addition, Fig.~\ref{fig:Butterflies} depicts how each parameter can independently drive the system through all topological phase transitions, irrespective of the other parameters. Furthermore, Fig.~\ref{fig:phase_diagram_comparison} seems to suggest that it is not a full picture of the phase relations.
To obtain a more complete view of the topology of the Sierpi\'nski gasket's gaps, we focus our attention onto two gaps: the previously discussed flux-induced gap around $\eta \approx 0.46$, closest to the half-filling considered in the original Haldane model, and a fractal gap around $\eta \approx 0.4$, corresponding to the largest gap induced by the fractality of the Sierpi\'nski gasket. The results are depicted in Fig.~\ref{fig:Fractal_Phase}. 

In Fig.~\ref{fig:Fractal_Phase}(a), the energy spectrum of a fifth-generation Sierpi\'nski gasket is shown. In this spectrum, the vertical lines represent the two values of $\eta$ of interest, colored in blue (flux-induced gap) and purple  (fractal gap). All states below these lines are filled and the last filled state is marked in the corresponding color.  In Fig.~\ref{fig:Fractal_Phase}(b), phase diagrams in terms of $\Phi$ and $M/t$ are shown for both of these gaps for six different values of the complex NNN hopping amplitude $\lambda$. Here, the background colors (blue and purple) indicate to which filling fraction $\eta$ the phase diagram belongs. The pink and green colors indicate the masked RSCN $\tilde{\nu}(P)$ of a fifth-generation Sierpi\'nski gasket with staggered mass $M$ and complex NNN hopping phase $\Phi$ for the corresponding filling fraction. 
As expected, when the NNN hopping strength $\lambda=0$, there are no topological phases, for both values of $\eta$. 
However, for $\lambda=0.2$ and $\eta \approx 0.46$ (blue), a wide range of topological states is observed, which are anti-symmetric around $\Phi =0$. 
As $\lambda$ is increased further, (inverted) conical shapes form around $\Phi=0$ and $\Phi = \pm \pi$. Furthermore, when $\lambda \geq 0.6$, two droplets start to emerge around $\Phi=\pm \pi/2$ and $M=0$, with opposite RSCN compared to their surroundings. 

For the largest fractal gap, which corresponds to $\eta \approx 0.4$ (purple), topological regions induced by $M$ and $\Phi$ are still observed, but now they are more restricted.
Increasing the value of $\lambda$ results in a rising level of complex behavior, reminiscent of Arnold's tongues \cite{adjanElevenPapersNumber1965}. Such Arnold tongues are also observed in the fractal seen in the parameter space of dynamical systems \cite{boylandBifurcationsCircleMaps1986}.
Here, interesting and complex behavior seems to grow from the center outwards on top of some conic shape. This conic shape reaches lower values of $M$ by increasing $\lambda$. At values of $\lambda$ between $0.8t$ and $t$, a droplet emerges at $\Phi = \pm \pi$ and $M=0$. 
At these filling fractions, the sinusoidal shape of the original Haldane model is completely lost, and the fractality of this system seems to induce much more intricate behavior. In Appendix~\ref{app:generations}, the influence of the chosen generation of the fractal is investigated. There, we present the phase diagram of a third, fourth, fifth, and sixth generation of the Sierpi\'nski gaskets with a strong complex NNN hopping, $\lambda = 0.8t$.

These results point to a very interesting possibility, as there are vertical lines of constant complex NNN hopping phase but variable staggered mass along which the topological marker goes from $1 \rightarrow 0 \rightarrow -1$, or vice versa. This could be useful because the complex NNN hopping phase and strength are material properties, but the Semenoff mass can be tuned by applying a perpendicular electric field \cite{ezawaTopologicalInsulatorHelical2012}, if the system has any buckling. In this case, 
\begin{equation*}
    \tilde{M} \rightarrow M + lE_\perp,
\end{equation*}
where $\tilde{M}$ is the new staggered mass, $l$ is the out-of-plane spatial separation between the sublattices, and $E_\perp$ the amplitude of the perpendicular electric field. 
This provides a very useful knob to experimentally tune the topological properties in a simple and direct way.

\begin{figure}
    \includegraphics{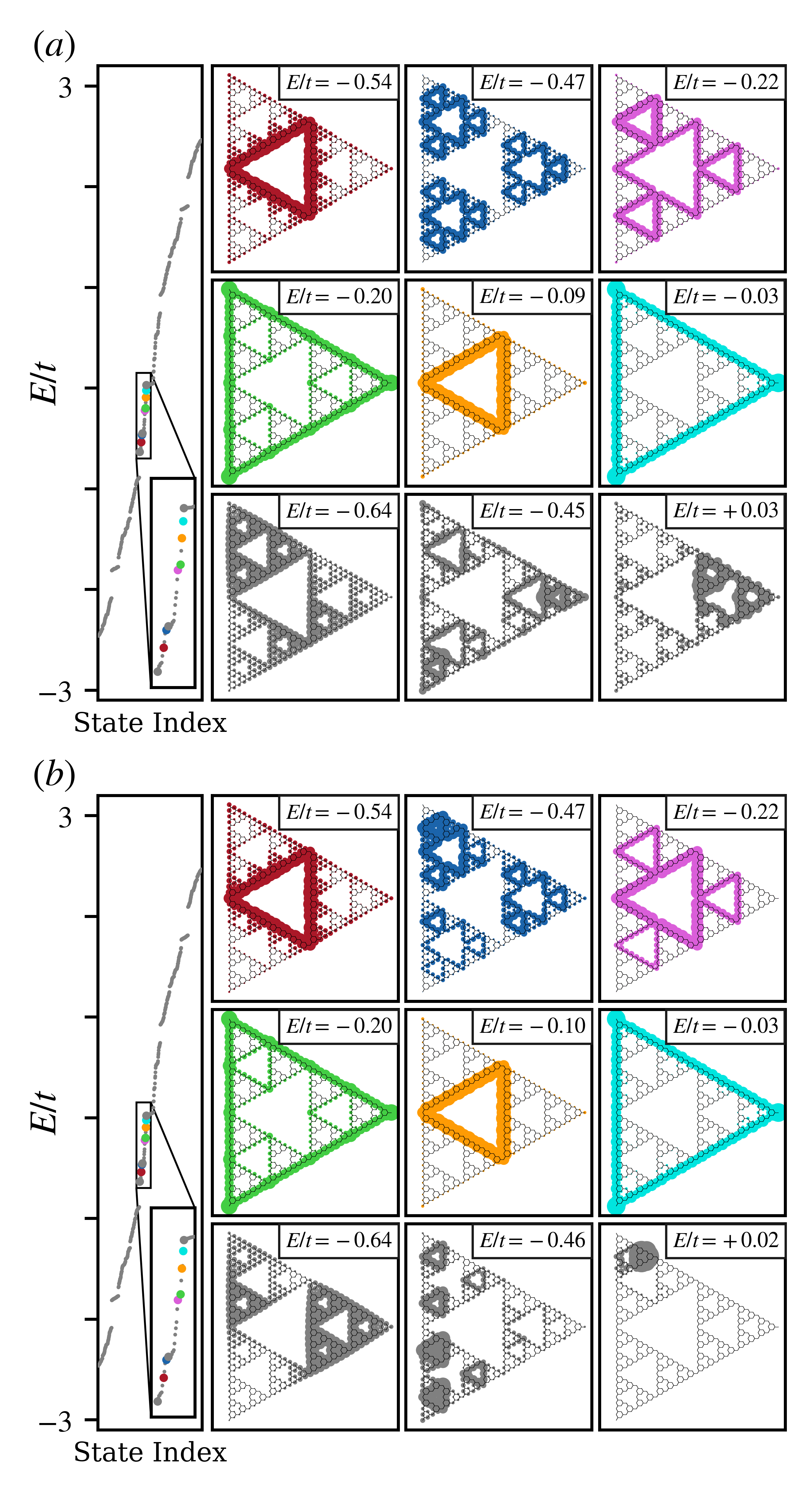}
    \caption{The effect of disorder on six edge states (colored) shown in Fig.~\ref{fig:TopoStates} and three additional bulk states (gray). Here, some arbitrary on-site disorder between $-\Delta$ and $\Delta$ is introduced. The disorder parameter $\Delta$ is $0.02\lambda$, and $0.2\lambda$ for (a) and (b), respectively. }
    \label{fig:disorder}
\end{figure}

\section{\label{sec:disorder} Effect of disorder on the topological states}
The robustness of topological states to disorder is one of the most interesting properties for applications. Here, we will introduce some arbitrary on-site disorder $\delta_i$ to probe the robustness of a set of states and verify whether these behave topologically. To model this on-site disorder, we add an on-site energy $\delta_i$ randomly sampled from the range $-\Delta$ to $\Delta$. The Hamiltonian with this disorder then reads
\begin{equation*}
    H = H_0 + \sum_i{\delta_i c^\dag_i c_i},
\end{equation*}
where $H_0$ is the original Hamiltonian Eq.~(\ref{eq:hamil}) and the second term describes the on-site disorder.

In Fig.~\ref{fig:disorder}, we show the energy spectra of and some states on the Sierpi\'nski gasket considered in Fig.~\ref{fig:TopoStates} with this disorder for two different values of $\Delta$. We depict the six edge-localized, topological states from Fig.~\ref{fig:TopoStates} (colored) together with three bulk-like states (gray). In Fig.~\ref{fig:disorder}(a), the disorder parameter $\Delta = 0.02 \lambda$, and in (b) $\Delta = 0.2 \lambda$. For the smaller disorder, the edge states are virtually unaffected. For a disorder of about 20\% of the NNN hopping, the topological states remain fully localized on the original edges, although the spatial symmetries are now clearly broken. This is similar to the behavior observed for the quantum spin Hall effect exhibited by bismuth fractals on InSb \cite{canyellasTopologicalEdgeCorner2024}. This behavior corroborates the topological nature of these edge-localized states (colored). In contrast, the bulk states (gray), change their localization significantly when the strength of the disorder is changed, because these states lack topological protection.

\section{\label{sec:conclusion}     Conclusions}
To conclude, the Haldane model on a single Sierpi\'nski gasket was considered. The energy spectra revealed the opening of flux-induced and fractal gaps.
A staggered mass, which breaks the symmetry between the two sublattices, combined with a complex NNN hopping was shown to cause the formation of a set of topological states in the Sierpi\'nski gasket. We investigated the robustness of these states against disorder and found that they remained fully edge-localized, indicative of their topological nature.
To further characterize the topology of these states, we calculated the real-space Chern number, as proposed by Kitaev \cite{kitaevAnyonsExactlySolved2006}. Using this method, we elucidated the relationship between the topological phases of the two types of gaps in these fractal structures and the different parameters of the model. 
Then, we compared the obtained phase diagram with previous results in the literature and showed that the Sierpi\'nski gasket did not fit the original phase diagram presented by Haldane in two dimensions \cite{haldaneModelQuantumHall1988} or the squeezed phase diagram of a doubled Sierpi\'nski gasket, as considered in Ref.~\cite{liFractalityinducedTopologicalPhase2023}. 
By thoroughly investigating the topological phases arising in the flux-induced gaps and the largest fractal gap, we found that more intricate structures arise. These structures are reminiscent of Arnold's tongues, suggestive of a possible connection to the fractality of this system. Our work lays a solid foundation for further studies of the enhancement of topological phases by fractality. 

In the future, it would be interesting to extend this analysis to other topological models and/or other fractal structures. Furthermore, it could be interesting to investigate which role self-similarity plays and to make predictions for the thermodynamic limit of this fractal. Lastly, it would be interesting to experimentally realize the proposed topological phase transitions by tuning solely an applied perpendicular electric field.
  
\begin{figure*}
	\includegraphics[]{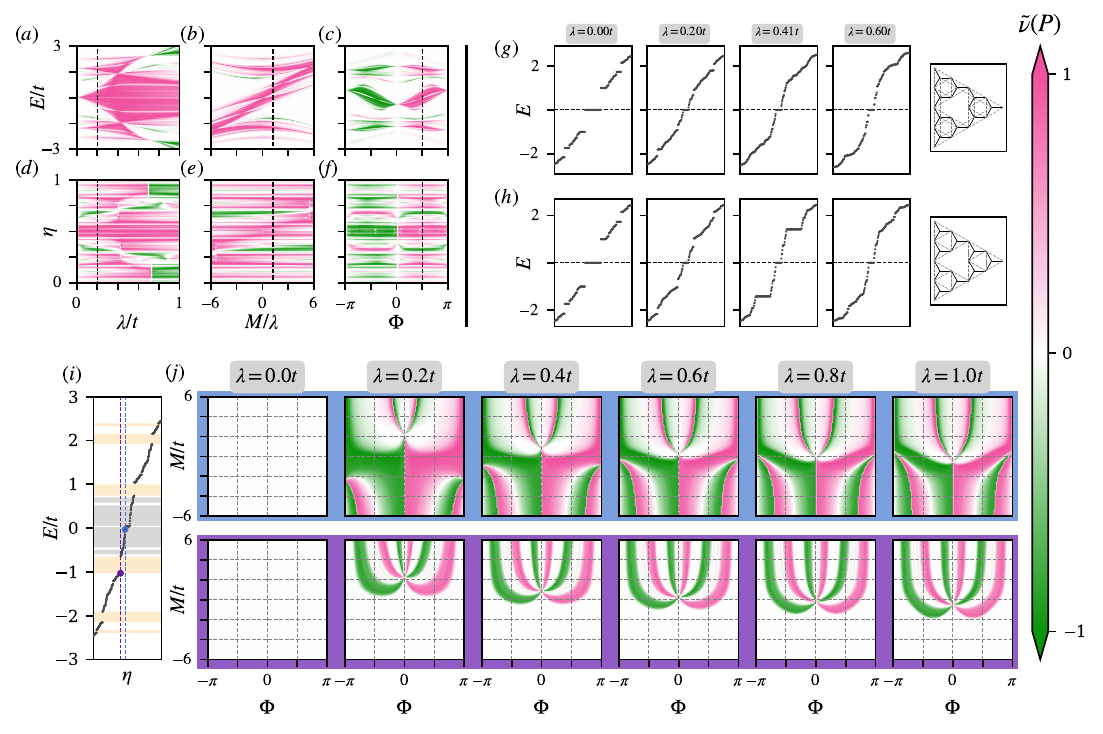}
	\caption{The results on a void-respecting Haldane model. (a)-(f) The Sierpi\'nski cocoons of this void respecting Haldane model, in terms of energy $E$ [\ref{fig:voidHaldane}(a)-(c)] and in terms of the filling fraction $\eta$ [\ref{fig:voidHaldane}(d)-(f)]. In (g) and (h), energy spectra for a range of $\lambda$'s are presented. (g) corresponds to the SOC-like fractal Haldane model, and (h) to the void-respecting fractal Haldane model, as shown in the accompanying lattice of a second-generation fractal of each of these models. In (i), we show an energy spectrum, with the filling fractions of interest marked, and the corresponding phase diagrams are presented on a backdrop of matching color in (j), akin to Fig.~\ref{fig:Fractal_Phase}.}
	\label{fig:voidHaldane}
\end{figure*}
      
\begin{acknowledgments}
The authors want to thank R. Arouca for a careful reading of the manuscript. 
Z.F.O., L.E., and C.M.S.~acknowledge the research program “Materials for the Quantum Age” (QuMat) for financial support. This program (registration number 024.005.006) is part of the Gravitation program financed by the Dutch Ministry of Education, Culture and Science (OCW). A.M. and C.M.S. acknowledge the TOPCORE project with project number OCENW.GROOT.2019.048 which
is financed by the Dutch Research Council (NWO)
\end{acknowledgments}

\appendix
\section{\label{app:voidHaldane} Results of the void-respecting fractal Haldane model}
In this appendix, we present the main results of this paper again, but now without the void-crossing complex NNN hoppings. 
In Fig.~\ref{fig:voidHaldane}(a)-(f), the Sierpi\'nski cocoons are shown for both energy $E$, and filling fraction $\eta$, in terms of the different parameters, analogous to their presentation in Fig.~\ref{fig:Butterflies}.
Most of these are quite similar to the results presented in the main text for a SOC-like Haldane model, but the $\tilde{\nu}(P) = 1$ phase is more dominant here. In Figs.~\ref{fig:voidHaldane}(a) and (d), the phase transitions from $\tilde{\nu}(P) = 1$ to $\tilde{\nu}(P) = -1$ are lost for most bands, except for a few of the gaps away from $E=0$.
In Figs.~\ref{fig:voidHaldane}(b) and (e), we also observe less gaps with $\tilde{\nu}(P) = -1$. Finally, in Figs.~\ref{fig:voidHaldane}(c) and (f) the topological phases are better separated in $\tilde{\nu}(P) = -1$ for $\Phi < 0$ and $ \tilde{\nu}(P) = +1$ for $\Phi > 0$.

In Figs.~\ref{fig:voidHaldane} (g) and (h), we present the energy spectra for different values of $\lambda$ (at $M = 0$) for the SOC-like Haldane model of the main text (g), and the void-respecting Haldane model (h). 
These spectra are accompanied by a visual representation of the second generation of the fractal model considered, in which the NN and NNN hoppings are depicted in black (full) and gray (dashed) lines, respectively. 
Here, we observe how the inclusion of the complex NNN hoppings across the voids influences the dispersion of the spectra.
Most notably, the emergence of a large flat band is observed for $\lambda = 1/\sqrt{6} \approx 0.41$.
\begin{figure*}
	\includegraphics[]{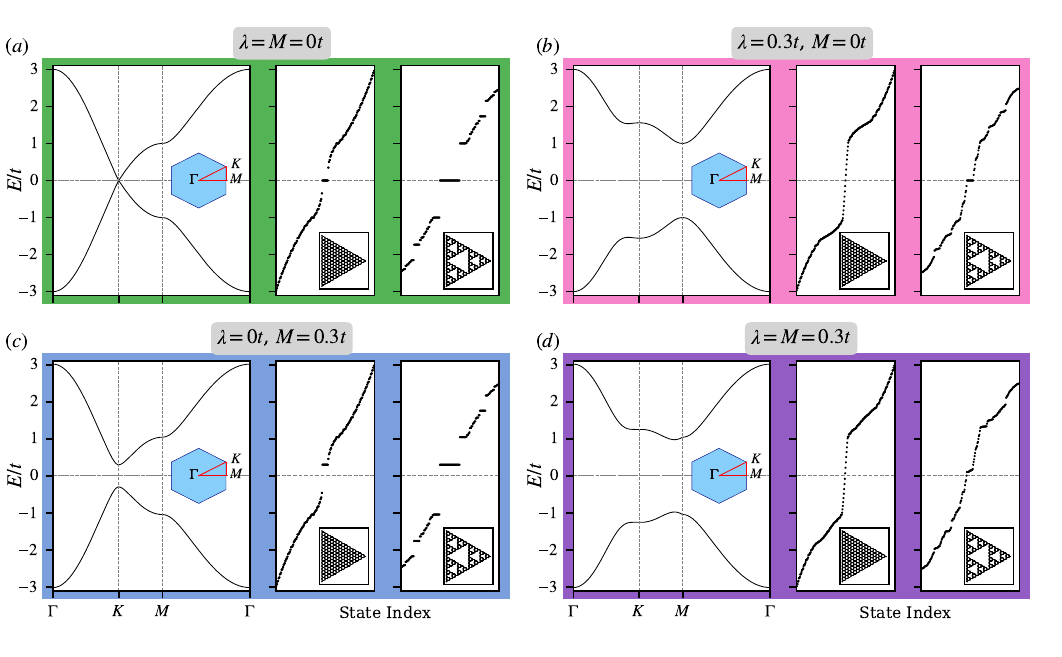}
	\caption{The energy spectra of the SOC-like Haldane model in the bulk, in a triangular flake, and in the Sierpi\'nski gasket. (a) The spectra for $\lambda=M=0t$, corresponding to Figs.~\ref{fig:Spectra}(a)-(c) of the main text, (b) the spectra for $\lambda=0.3t$ and $M=0$, (c) for $\lambda=0t$ and $M=0.3t$, and (d) for $\lambda=M=0.3t$. This shows that either a nonzero Semenoff mass $M$ or a nonzero complex NNN hopping can open a bulk gap, but that the topological nature of these gaps differs. Furthermore, one may observe that the symmetry of the spectrum is lifted for the finite-size models for $M>0$, and how $\lambda>0$ induces the emergence of states crossing the gaps of the respective models.}
	\label{fig:AP_HaldaneTerms}
\end{figure*}

In Fig.~\ref{fig:voidHaldane}(j), we present a set of phase diagrams of the system shown in Fig.~\ref{fig:voidHaldane}(h), for different values of $\lambda$ and two different filling fractions $\eta = \{0.46, 0.4\}$, analogous to Fig.~\ref{fig:Fractal_Phase}. Here, we observe that the regions corresponding to a topological phase are both more stable and better quantized, but no increasingly complex patterns are observed. 
For the flux-induced gap (blue), we observe a similar trend as for the SOC-like Haldane model, where two (inverted) conical shapes form around $\Phi = 0$ and $\Phi=\pm \pi$, reaching larger values of $\abs{M}$ with increasing $\lambda$. In addition, distinct upward (downward) phase boundaries (white curves) are shown in the center of these conical shapes.
For the fractal gap (purple), a similar conic shape arises, which reaches lower values of $M$ as $\lambda$ is increased, but no growing complexity is observed.
Finally, in contrast with the SOC-like Haldane model, no droplet shapes emerge in the phase diagrams of the considered gaps of the void-respecting Haldane model.

\section{\label{app:individualTerms} Effect of the individual Haldane terms}
Here, we present the separate effect of the staggered mass and complex NNN hopping of the SOC-like Haldane model. In Fig.~\ref{fig:AP_HaldaneTerms}, the energy spectra of the Haldane model on a bulk, a triangular flake and the Sierpi\'nski gasket are depicted for different combinations of the staggered mass $M$, and the complex NNN hopping strength $\lambda$. In Fig.~\ref{fig:AP_HaldaneTerms}(a), no Semenoff mass or complex NNN hopping is considered. We note the lack of a gap in both the spectra of the bulk and of the triangular flake, but also the emergence of fractal gaps, as discussed in the main text.
In Fig.~\ref{fig:AP_HaldaneTerms}(b), only a strong complex NNN hopping is taken into account. This opens a large gap in the bulk spectrum, and the system is in a topological phase because $3\sqrt{3}\lambda \sin{\Phi} > \abs{M}$. In the triangular flake, this gap is populated by topological states, as expected from the bulk-boundary correspondence. In the fractal, a complex NNN hopping of this strength fully populates the largest fractal gap with states.
\begin{figure*}
	\includegraphics[width=\textwidth]{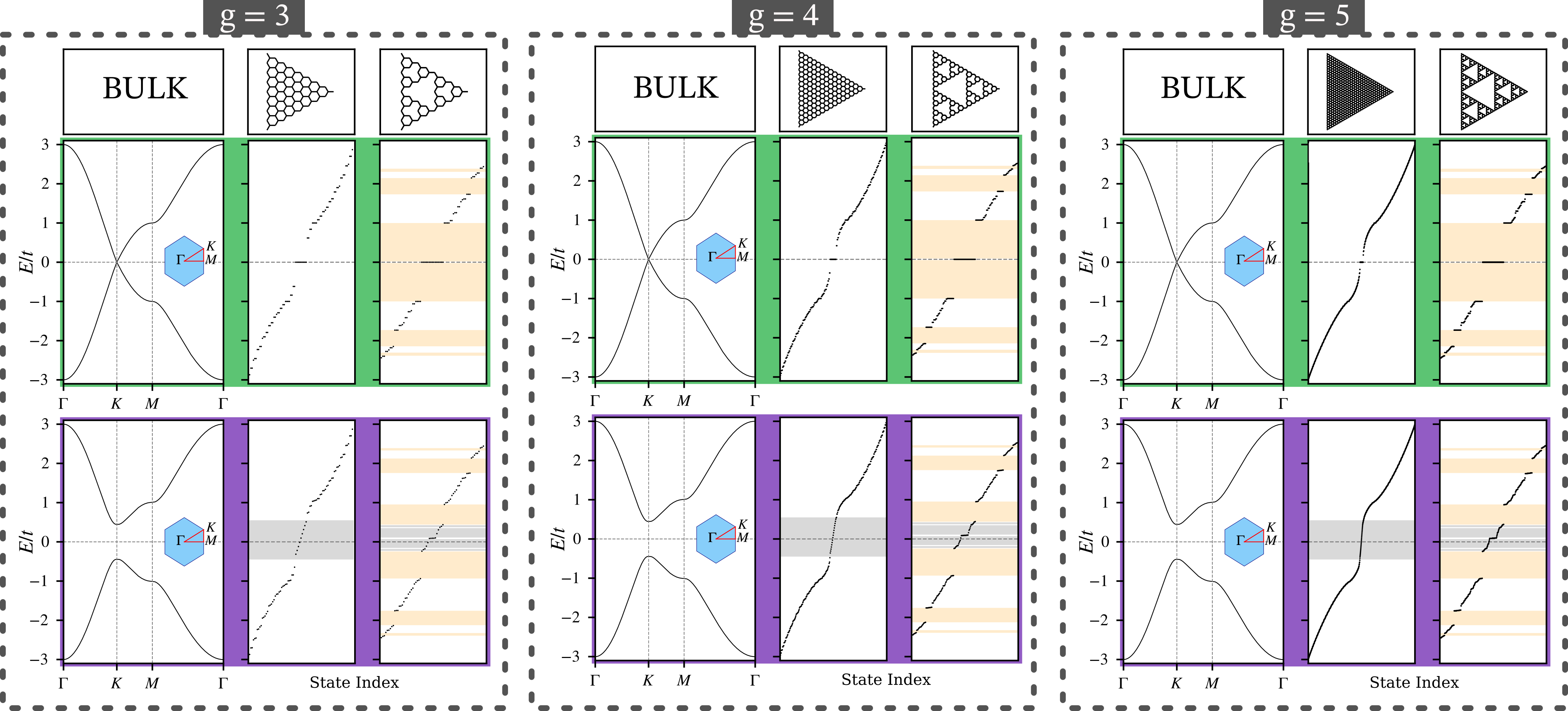}
	\caption{The energy spectra for different sizes of the triangular flake and the Sierpi\'nski gasket, each analogous to Fig.\ref{fig:Spectra}. The sizes are chosen to correspond to a generation $g=3,4,5$, respectively. The fractal gaps are marked in orange and the flux-induced gaps are marked in gray. These results allow us to distinguish between intrinsic and finite-size effects, such as the spurious gap opening in the spectrum of a triangular flake. Notably, the size and position of the fractal gaps do not depend on the generation, indicating that the gaps remain in the thermodynamic limit.}
	\label{fig:finiteSize}
\end{figure*}
In Fig.~\ref{fig:AP_HaldaneTerms}(c), only a non-zero staggered mass is taken into account, $M=0.3t$. Now, the system is in a trivial phase, and a bulk band gap is observed. As expected, there are no in-gap states for the triangular flake. In addition, we observe that the middle flat band is shifted from $E=0$ to $E=M$. Similarly, the gaps of the Sierpi\'nski gasket remain open, and the middle flat band is shifted to $E=M$. This raising of the energy of the central flat band of the OBC geometries breaks the spectral symmetry of these models. This is a consequence of $M\neq 0$, in combination with the asymmetry in the number of lattice sites in each sublattice.
In Fig.~\ref{fig:AP_HaldaneTerms}(d), these effects are combined, and both the ``in-gap'' states and the broken symmetry can be observed.

\section{\label{app:finiteSize} Investigation of finite-size effects}
\begin{figure*}
	\includegraphics[]{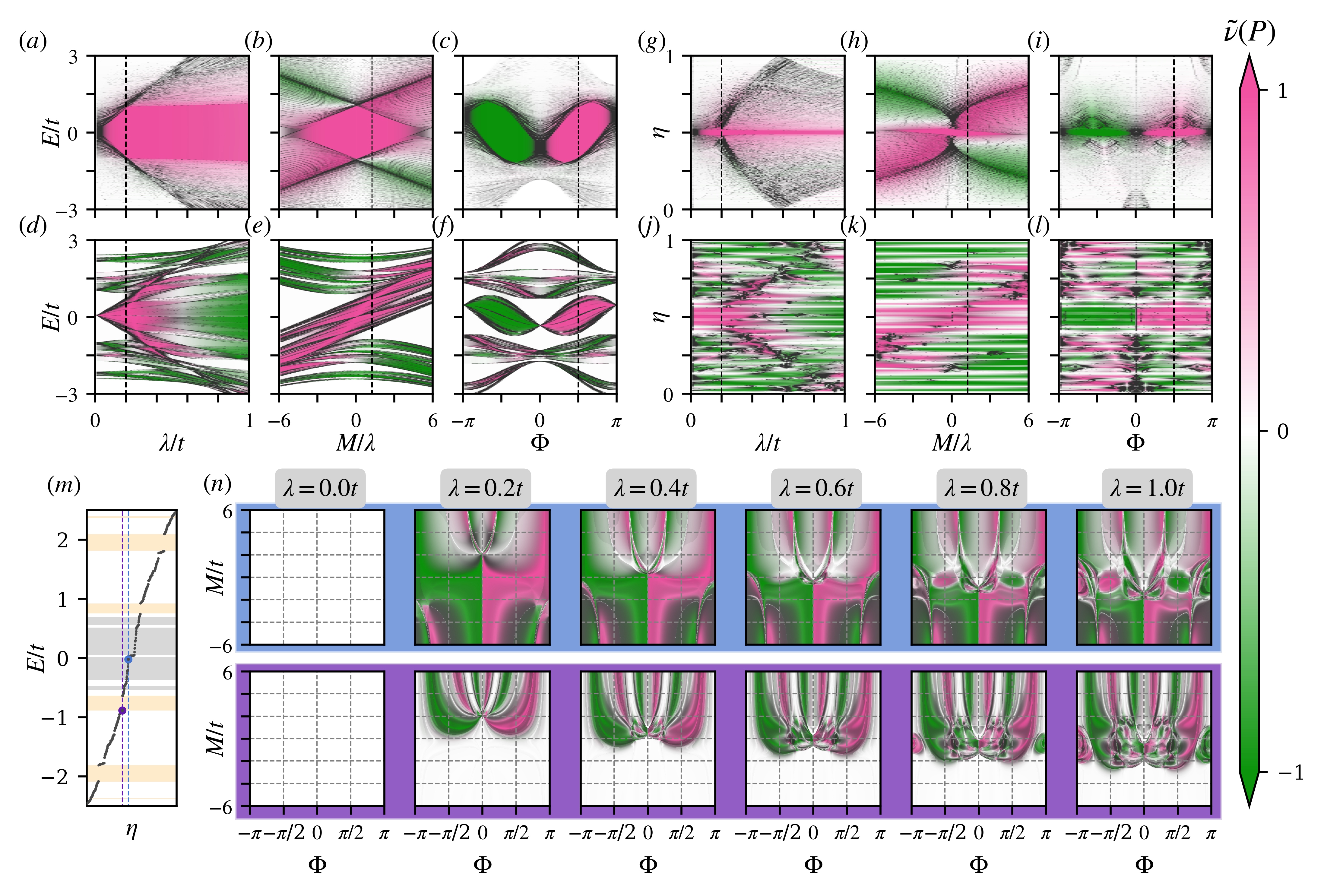}
	\caption{The same results as presented in the main text in Figs.~\ref{fig:Butterflies} and \ref{fig:Fractal_Phase}. However, the transparency mask is now replaced by a black mask to indicate the regions of unstable RSCN. As expected, most topological phases are bounded by unstable regions of the RSCN.}
	\label{fig:maskvisual}
\end{figure*}
Here, we investigate how the energy spectra discussed in Sec.~\ref{sec:energy} scale with the size of the system. To this end, we have considered structures of sizes corresponding to generations $g = 3,4, 5$ in Fig.~\ref{fig:finiteSize}.
For the smallest systems ($g=3$), a apparent gap around $E=0$ can be seen in the spectrum of a triangular flake. However, when we compare this spectrum to the spectrum of triangular flakes with more sites ($g = 4$), the gap shrinks. For the largest considered system ($g=5$), this apparent gap has shrunk even further. 
Therefore, we conclude that this gap opening is a finite-size effect. This can be understood by a simple analogy with the particle-in-a-box model. In this model, the wavefunction must have nodes on the boundaries of the system. Consequently, the wavefunction with the lowest allowed frequency has frequency $f \approx (2L)^{-1}$, with $L$ a characteristic length scale. 
Therefore, the size of the triangular flake puts a lower limit on the allowed non-zero energies because the low-energy modes correspond to low-frequency wavefunctions.

It is of note that the fractal gaps behave very differently. The size and position of these gaps does not change when comparing different generations. Therefore, we conclude that these gaps are not a finite-size effect, but truly an intrinsic property caused by the fractality of the system.

\section{\label{app:visualMask}Visualization of the RSCN mask}
In the main text, we discussed how the value of the RSCN is not enough to conclusively decide about the non-triviality of a state. We emphasized that one also needs to establish whether such a state belongs to a ``bulk band'' or if such a state is ``in-gap''. In this work, we took a numerical derivative and mapped the results to a value between $0$ and $1$ to form a transparency mask by taking an exponential,
\begin{equation*}
    \mathrm{Transparency}(\nu(P)) = 2^{-\nu'(P)/ \nu'_{1/2}}.
\end{equation*}
Here, $\nu'(P)$ is the numerical derivative of $\nu$ and $\nu'_{1/2} = 0.05$ is the value of this derivative at which the transparency is one-half.

In Fig.~\ref{fig:maskvisual}, we follow a similar procedure, but now the background of each figure is black, such that this mask actually mixes in a black color. In this manner, not only the value but also the stability of the RSCN is depicted. Interestingly, while in most graphs just the boundaries of the topological phase are affected, in Fig.~\ref{fig:maskvisual}(b), four bright black and straight lines arise. These lines correspond to the bump in Fig.~\ref{fig:RSCN}, which was discussed in the main text. This seems to suggest that there is some linear relationship between the strength of the staggered mass and the energy corresponding to this unstable divergence of the RSCN. This calls for further investigation, with alternative methods.

\section{\label{app:generations} Phase diagram for different generations}
\begin{figure*}
    \includegraphics{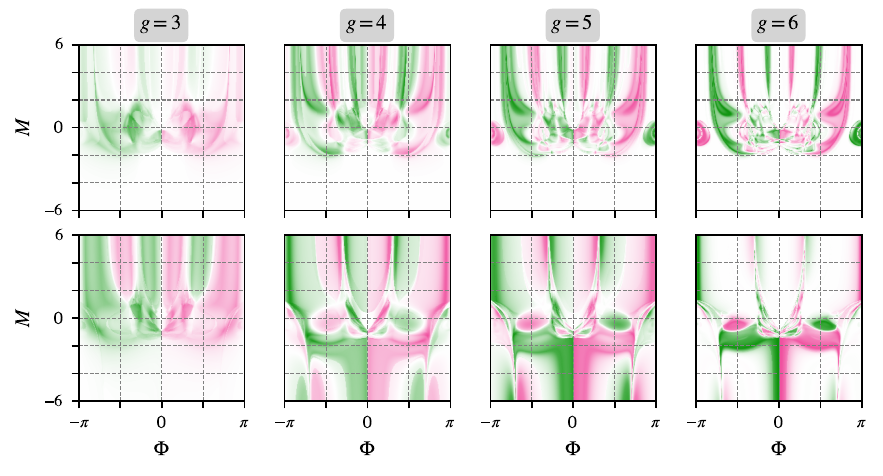}
    \caption{The phase diagrams of both a flux-induced and a fractal gap for different generations $g$ of the Sierpi\'nski gasket, with a complex NNN hopping strength $\lambda = 0.8t$. In (a) a third, (b) a fourth, (c) a fifth, and (d) a sixth generation fractal are considered. The phase diagrams become more complex as the generation increases.}
    \label{fig:gen}
\end{figure*}
In this appendix, we illustrate how the generation of the considered fractal influences the phase diagram. As the generation increases, the fractal consists of more and more sites, and the structure becomes a better approximation of the ``true'' (generation to infinity) Sierpi\'nski gasket. In addition, the increased number of sites also allows for choosing the subsets $A$, $B$, and $C$ larger. In this case, the RSCN becomes a better approximation of the bulk Chern number.

In Fig.~\ref{fig:gen}, the phase diagrams of both a flux-induced gap and the largest fractal gap of the Sierpi\'nski gasket are shown. Here, the generation $g$ ranges from 3 to 6 and the complex NNN hopping strength $\lambda = 0.8t$. In Fig.~\ref{fig:gen}(a), the phase diagrams of the third-generation fractal are shown. Most notably, we see that the RSCN is poorly quantized and little patterns can be distinguished. This is because each subset $A$, $B$, and $C$ now contains about four sites, which is not large enough for the RSCN to confidently predict the topology of a gap.

Similarly, in Figs.~\ref{fig:gen}(b), (c), and (d) the phase diagrams are shown for a fourth, fifth and sixth generation of the fractal, respectively. Here, we recognize two patterns: firstly, as the generation increases, the RSCN is better quantized to integer values, and secondly, with increasing generation the structure becomes more complex.

This first pattern is best observed for larger values of $\abs{M}$ in both phase diagrams, where the transition from $\abs{\tilde{\nu}(P)}=1$ to zero and back becomes more stepwise, and the gradual transitions of the lower generations is slowly lost, caused by the better quantization of the RSCN.

The second pattern is best observed around $\Phi = M = 0$ in the phase diagrams of both gaps, and in the droplet around $\Phi = \pm \pi$ and $M=0$ of the fractal gap. Here, we observe how even more features appear in the phase diagram. The individual topological regions shrink, but one can identify finer topological structure, reminiscent of the self-similarity of the fractal. Because we have only changed the generation of the fractal,
this suggests that these patterns are indeed driven by the fractality of the structure.

\bibliography{bibfile}

\end{document}